\documentclass[11pt,graphicx,subfigure,axodraw]{article}
\usepackage{amsfonts}
\usepackage{amssymb}
\usepackage{amsmath,amsfonts,amssymb}
\usepackage{epstopdf}
\usepackage[section]{placeins}
\usepackage{hyperref}
\usepackage{booktabs}
\usepackage{cleveref}
\crefname{equation}{Eq.}{Eqs.}
\crefname{figure}{Fig.}{Figs.}
\crefname{table}{Table}{Tables}
\crefname{section}{Section}{Sections}

\usepackage[usenames,dvipsnames]{color}
\usepackage{graphicx}
\usepackage{subfigure}
\usepackage{float}
\def\rmuu{\gamma^{\mu}}
\def\rmud{\gamma_{\mu}}
\def\PL{{1-\gamma_5\over 2}}
\def\PR{{1+\gamma_5\over 2}}
\def\sinW2{\sin^2\theta_W}
\def\AEM{\alpha_{EM}}
\def\mul{M_{\tilde{u} L}^2}
\def\mur{M_{\tilde{u} R}^2}
\def\mdl{M_{\tilde{d} L}^2}
\def\mdr{M_{\tilde{d} R}^2}
\def\mz2{M_{z}^2}
\def\c2b{\cos 2\beta}
\def\au{A_u}
\def\ad{A_d}
\def\cob{\cot \beta}
\def\v#1{v_#1}
\def\tb{\tan\beta}
\def\epem{$e^+e^-$}
\def\KK{$K^0$-$\overline{K^0}$}
\def\wi{\omega_i}
\def\xj{\chi_j}
\def\Wmu{W_\mu}
\def\Wnu{W_\nu}
\def\m#1{{\tilde m}_#1}
\def\mH{m_H}
\def\mw#1{{\tilde m}_{\omega #1}}
\def\mx#1{{\tilde m}_{\chi^{0}_#1}}
\def\mc#1{{\tilde m}_{\chi^{+}_#1}}
\def\mwi{{\tilde m}_{\omega i}}
\def\mxi{{\tilde m}_{\chi^{0}_i}}
\def\mci{{\tilde m}_{\chi^{+}_i}}

\def\ch{{\tilde\chi^{+}_1}}
\def\c2{{\tilde\chi^{+}_2}}

\def\tt{{\tilde\theta}}

\def\tp{{\tilde\phi}}

\def\mz{M_z}
\def\sw{\sin\theta_W}
\def\cw{\cos\theta_W}
\def\cb{\cos\beta}
\def\sb{\sin\beta}
\def\rwi{r_{\omega i}}
\def\rxj{r_{\chi j}}
\def\rfp{r_f'}
\def\Kik{K_{ik}}
\def\Fq2{F_{2}(q^2)}
\def\f{\({\cal F}\)}
\def\d1{{\f(\tilde c;\tilde s;\tilde W)+ \f(\tilde c;\tilde \mu;\tilde W)}}
\def\tw{\tan\theta_W}
\def\sec2w{sec^2\theta_W}
\begin{document}
\baselineskip 18pt
\def\today{\ifcase\month\or
 January\or February\or March\or April\or May\or June\or
 July\or August\or September\or October\or November\or December\fi
 \space\number\day, \number\year}
\def\thebibliography#1{\section*{References\markboth
 {References}{References}}\list
 {[\arabic{enumi}]}{\settowidth\labelwidth{[#1]}
 \leftmargin\labelwidth
 \advance\leftmargin\labelsep
 \usecounter{enumi}}
 \def\newblock{\hskip .11em plus .33em minus .07em}
 \sloppy
 \sfcode`\.=1000\relax}
\let\endthebibliography=\endlist
\def\lsim{\ ^<\llap{$_\sim$}\ }
\def\gsim{\ ^>\llap{$_\sim$}\ }
\def\r2{\sqrt 2}
\def\beq{\begin{equation}}
\def\eeq{\end{equation}}
\def\beqn{\begin{eqnarray}}
\def\eeqn{\end{eqnarray}}
\def\rmuu{\gamma^{\mu}}
\def\rmud{\gamma_{\mu}}
\def\PL{{1-\gamma_5\over 2}}
\def\PR{{1+\gamma_5\over 2}}
\def\sinW2{\sin^2\theta_W}
\def\AEM{\alpha_{EM}}
\def\mul{M_{\tilde{u} L}^2}
\def\mur{M_{\tilde{u} R}^2}
\def\mdl{M_{\tilde{d} L}^2}
\def\mdr{M_{\tilde{d} R}^2}
\def\mz2{M_{z}^2}
\def\c2b{\cos 2\beta}
\def\au{A_u}
\def\ad{A_d}
\def\cob{\cot \beta}
\def\v#1{v_#1}
\def\tb{\tan\beta}
\def\epem{$e^+e^-$}
\def\KK{$K^0$-$\bar{K^0}$}
\def\wi{\omega_i}
\def\xj{\chi_j}
\def\Wmu{W_\mu}
\def\Wnu{W_\nu}
\def\m#1{{\tilde m}_#1}
\def\mH{m_H}
\def\mw#1{{\tilde m}_{\omega #1}}
\def\mx#1{{\tilde m}_{\chi^{0}_#1}}
\def\mc#1{{\tilde m}_{\chi^{+}_#1}}
\def\mwi{{\tilde m}_{\omega i}}
\def\mxi{{\tilde m}_{\chi^{0}_i}}
\def\mci{{\tilde m}_{\chi^{+}_i}}
\def\mz{M_z}
\def\sw{\sin\theta_W}
\def\cw{\cos\theta_W}
\def\cb{\cos\beta}
\def\sb{\sin\beta}
\def\rwi{r_{\omega i}}
\def\rxj{r_{\chi j}}
\def\rfp{r_f'}
\def\Kik{K_{ik}}
\def\Fq2{F_{2}(q^2)}
\def\f{\({\cal F}\)}
\def\d1{{\f(\tilde c;\tilde s;\tilde W)+ \f(\tilde c;\tilde \mu;\tilde W)}}
\def\tw{\tan\theta_W}
\def\sec2w{sec^2\theta_W}
\def\ch{{\tilde\chi^{+}_1}}
\def\c2{{\tilde\chi^{+}_2}}

\def\tt{{\tilde\theta}}

\def\tp{{\tilde\phi}}

\def\mz{M_z}
\def\sw{\sin\theta_W}
\def\cw{\cos\theta_W}
\def\cb{\cos\beta}
\def\sb{\sin\beta}
\def\rwi{r_{\omega i}}
\def\rxj{r_{\chi j}}
\def\rfp{r_f'}
\def\Kik{K_{ik}}
\def\Fq2{F_{2}(q^2)}
\def\f{\({\cal F}\)}
\def\d1{{\f(\tilde c;\tilde s;\tilde W)+ \f(\tilde c;\tilde \mu;\tilde W)}}

\def\b{$\cal{B}(\tau\to\mu \gamma)$~}


\def\tw{\tan\theta_W}
\def\sec2w{sec^2\theta_W}
\newcommand{\pxn}[1]{{\color{red}{#1}}}
\newcommand{\ai}[1]{{\color{blue}{#1}}}

\begin{titlepage}
\begin{center}
{\large {\bf
Electron EDM as a Sensitive Probe of PeV Scale Physics}}\\

\vskip 0.5 true cm
 Tarek  Ibrahim$^{a,b}$\footnote{Email:
  tibrahim@zewailcity.edu.eg},
Ahmad Itani$^{c}$\footnote{Email: a.itanis@bau.edu.lb},
  and Pran Nath$^{d}$\footnote{Emal: nath@neu.edu}
\vskip 0.5 true cm
\end{center}

\noindent
{$^{a}$ Department of  Physics, Faculty of Science,
University of Alexandria, Alexandria 21511, Egypt\footnote{Permanent address.} }\\
{$^{b}$ Center for Fundamental Physics, Zewail City of Science and Technology, Giza 12588, Egypt\footnote{Current address.}} \\
{$^{c}$
 Department of Physics, Faculty of Sciences, Beirut Arab University,
Beirut 11 - 5020, Lebanon} \\
{$^{d}$ Department of Physics, Northeastern University, Boston, Massachusetts  02115-5000, USA} \\


\centerline{\bf Abstract}
We give a quantitative analysis of the electric dipole moments 
as a probe of high scale physics. 
We focus on the electric dipole moment of the electron since the limit on it is the most stringent.
Further, theoretical computations of it are free of QCD uncertainties. 
 The analysis presented here first explores the probe of high scales via electron 
 electric dipole moment (EDM) within MSSM
 where the contributions to the EDM  arise from the chargino and the neutralino exchanges
 in loops.  Here it is shown that the  electron EDM 
 can probe mass scales from tens of TeV into the  PeV range.  The analysis  is then extended to 
 include a vectorlike generation which can mix with the three ordinary generations. Here new 
 CP phases arise and it is shown that the electron EDM now has not only a supersymmetric contribution 
 from the exchange of charginos and neutralinos but also a non-supersymmetric contribution 
 from the exchange of $W$ and $Z$ bosons.  It is further shown that the interference of the 
 supersymmetric and the non-supersymmetric contribution leads to the remarkable phenomenon
 where the electron EDM as a function of the slepton mass first falls and become vanishingly small
 and then rises again as the slepton mass increases This phenomenon arises as a consequence
 of cancellation between the SUSY and the non-SUSY contribution at low scales while at high
 scales the SUSY contribution dies out and the EDM is controlled by the non-SUSY contribution alone.
  The high mass scales that can be probed by the EDM are far in excess of what accelerators will be able to probe. 
  The sensitivity of the EDM to CP phases both in the SUSY and the non-SUSY sectors are also 
  discussed.

 \noindent
{\scriptsize
}
\medskip

\end{titlepage}
\section{Introduction \label{sec1}}
In the standard model the electric dipole moments of elementary particles are very small\cite{sm}. 
Thus for  the electron it is estimated that $|d_e|\simeq  10^{-38}$ $e$cm
and for the neutron the value ranges from $10^{-31}-10^{-33}$ $e$cm.
This is far beyond the current sensitivity of experiments to measure. 
However, in models of physics beyond the standard model much larger electric dipole moments,
orders of magnitude larger than those in the standard model,  can be obtained (for a review see~\cite{Ibrahim:2007fb}). 
Thus in the supersymmetric models the electric dipole moments of elementary particles such
as the electron and the quarks can be large enough that the current experimental upper limits
act as constraints on models. Indeed often in supersymmetric theories for light scalars, the 
electric dipole moments can lie in the region larger than the current upper limits for the 
electron and the neutron EDMs. This phenomenon is 
often referred to as the SUSY EDM problem. 
One solution to the SUSY EDM problem is the possibility that the CP phases
are small~\cite{earlywork}. Other possibilities allow for large,  even maximal, phases and the EDM
is suppressed via the sparticle masses being large~\cite{Nath:1991dn}
 or by invoking the so called cancellation mechanism~\cite{cancellation}
 where contributions 
from various diagrams that generate the electric dipole moment interfere destructively 
to reduce the electric dipole moment to a level below its experimental upper limit.\\

In the post Higgs boson discovery era the apparent  SUSY EDM problem can be turned around to ones
advantage as a tool to investigate high scale physics.  The logic of this approach is the following:
The high mass of the Higgs boson at 126 GeV requires a large loop correction to lift its value
from the tree level, which lies below the $Z$ -boson mass, up to the experimental value.   A large loop
correction requires that the scalar masses that enter in the Higgs boson loop be large so as to generate
the desired large correction which requires a high scale for the sfermion masses. 
Large sfermions masses help with suppression of flavor changing neutral currents.  They 
also help resolve the SUSY EDM problem
and  help stabilize the proton against decay via baryon and lepton number violating dimension five
operators in supersymmetric grand unified theories. \\

In this work we investigate the possibility that EDMs can be used as probes of high scale physics
as suggested in ~\cite{McKeen:2013dma,Moroi:2013sfa,Altmannshofer:2013lfa,Dhuria:2013ida}.
Specifically we focus here on the EDM of the  electron since it is by far the most 
sensitively determined one than any of the other EDMs. 
Thus the ACME Collaboration~\cite{Baron:2013eja} using the polar molecule thorium monoxide (ThO) measures the 
electron EDM so that 
\beqn
d_e =(-2.1 \pm 3.7_{\rm stat} \pm 2.5_{\rm syst}) \times  10^{-29} e{\rm cm}.
\label{1.1}
\eeqn
The above corresponds to an upper limit of 
\beqn
 |d_e| < 8.7\times  10^{-29} ~e{\rm cm}\ ,
 \label{1.2}
\eeqn
at 90\% CL. The corresponding upper limits on the EDM of the muon and  on the tau lepton are 
~\cite{pdg}
\beqn
 |d_\mu| < 1.9 \times 10^{-19} ~e{\rm cm}\ , \\
 \label{1.3}
  |d_\tau| <  10^{-17} ~e{\rm cm}\ ,
   \label{1.4}
   \eeqn
   and are not as stringent as the result of Eq. (\ref{1.2}) even after scaling in lepton mass is taken into account. 
   Further, the limit on $d_e$    is likely 
to improve by an order of magnitude or more in the future as the projections below indicate

\begin{align}
\label{1.5}
{\text Fr }\cite{Sakemi:2011zz}
      & ~~~~|d_e| \lesssim 1 \times 10^{-29} e{\rm cm}
      \\
      \label{1.6}
      {\rm YbF ~molecule} \cite{Kara:2012ay}
      &~~~ |d_e| \lesssim 1 \times 10^{-30} e{\rm cm}
      \\
    {\rm   WN ~ion} \cite{Kawall:2011zz}
      &~~~ |d_e| \lesssim 1 \times 10^{-30} e{\rm cm}
      \label{1.7}
\end{align}
In the analysis here we will first consider the case of MSSM where the CP phases enter in the 
soft parameters such as in the  masses $M_i$ (i=1,2) of the electroweak gauginos,
 and in the trilinear couplings $A_k$ and in the Higgs mixing parameter $\mu$. 
 Here we will investigate the scale of the slepton masses needed to reduce the electron EDM 
below its upper limit for the case when the CP phases are naturally ${\cal O}(1)$. We will
see that this scale will be typically high lying in the range of tens of TeV to a PeV
(For a discussion of PeV scale in the context of supersymmetry in previous works see, e.g., \cite{Wells:2004di}).
We will carry out  the analysis for the case where we extend MSSM to include a vector like leptonic 
multiplet and allow for mixings between the vector like multiplet and the three sequential generations.
We will  study the parametric dependence of the EDM  on the scalar masses, on
fermion masses of the vector like generation, on CP phases and on $\tan\beta$.\\

The outline of the rest of the paper is as follows: In \cref{sec2} we  discuss 
 the EDM of the electron within MSSM as a probe of the slepton masses.
In \cref{sec3} we extend the analysis of \cref{sec2} to MSSM with inclusion of a vector 
like leptonic  multiplet which allows for mixing between the vector multiplet and the three
sequential generations.  Here we give analytic results for the electron EDM arising from the 
supersymmetric exchange involving the chargino and neutralinos in the loops. 
We also compute the non-supersymmtric contributions involving the $W$ and the $Z$ exchange. 
In \cref{sec4} we give a numerical analysis of the limits on the mass scales that can be 
{accessed} using the results of \cref{sec3}. Conclusions are given in \cref{sec5}.
Further details of the MSSM model with a vector multiplet used in the analysis of 
\cref{sec3} are given in Appendices A-C.

\begin{figure}[t]
\begin{center}
{\rotatebox{0}{\resizebox*{14cm}{!}{\includegraphics{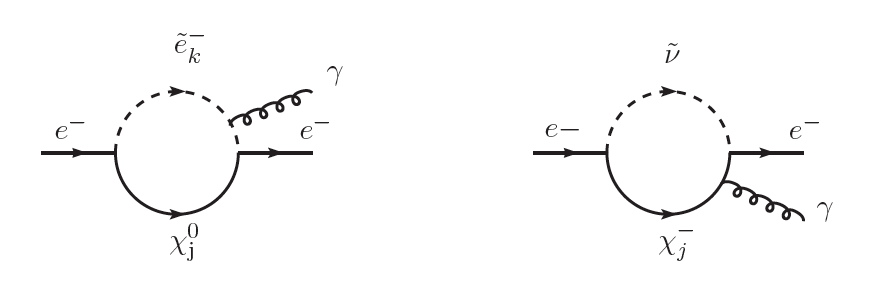}}\hglue5mm}}
\caption{The neutralino-slepton  exchange diagram (left) and the chargino -sneutrino exchange diagram (right) that contribute
to the  electric dipole moment of the electron in MSSM.}
 \label{fig1}
\end{center}
\end{figure}

\section{Probe of slepton masses in MSSM from the electron EDM constraint  \label{sec2}}
\begin{figure}[t]
\begin{center}
{\rotatebox{0}{\resizebox*{7.5cm}{!}{\includegraphics{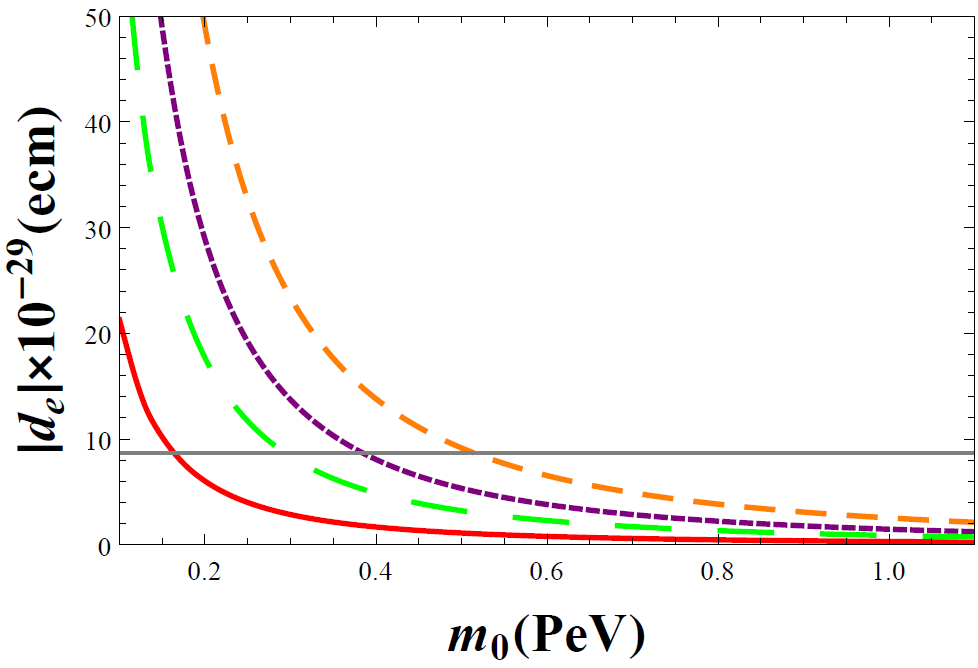}}\hglue5mm}}
{\rotatebox{0}{\resizebox*{7.5cm}{!}{\includegraphics{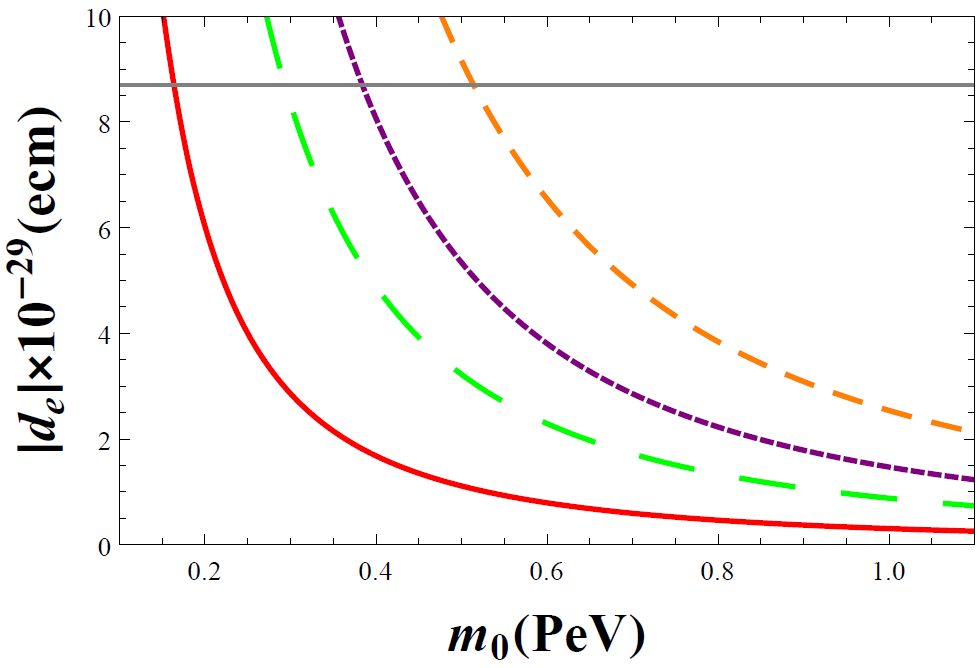}}\hglue5mm}}\\
\caption{{Left panel:}
A display of the electron EDM as a function of $m_0$ (where $m_0= M_{\tilde eL}= M_{\tilde e}$)
for different $\alpha_\mu$ (the phase of the 
Higgs mixing parameter $\mu$) 
 with the  mixings of the vector like generation with the regular three generations   set to zero. The curves are for 
 the cases $\alpha_\mu= -3$ (small-dashed, red), $\alpha_\mu=-0.5$ (solid), $\alpha_\mu=1$ (medium-dashed, orange), and $\alpha_\mu= 2.5$ (long-dashed, green). The horizontal solid line is the current upper limit on the electron EDM set at $|d_e|=8.7 \times 10^{-29}$. The other parameters are  $\text{$|\mu |$ = }4.1\times 10^2\text{ ,  $|$}M_1\text{$|$ = }2.8\times 10^2\text{ ,  $|$}M_2\text{$|$ = }3.4\times 10^2\text{ ,  $|$}A_e\text{$|$ = }3\times 10^6\text{ ,  }m_0^{\tilde{\nu}}\text{ = }4\times 10^6\text{ ,  $|$}A_0^{\tilde{\nu}}\text{$|$ = }5\times 10^6\text{ , tan$\beta $ = }30$ . All masses are in GeV,  phases in rad and EDM in $e$cm.The analysis shows that improvements in the electron EDM constraint can probe scalar masses in the 100 TeV- 1 PeV region and beyond. 
 {Right panel: The same as the left panel except that the region below the current experiment limit is blown
 up. The analysis shows that an improvement by a factor of ten can allow one to probe up to and beyond 1 PeV
 in mass scales.} 
 }
\label{fig2}
\end{center}
\end{figure}

The supersymmetric Feynman diagrams that contribute  to the electric dipole moment of the electron
involve the chargino-sneutrino exchange and the neutralino-slepton exchange as shown in \cref{fig1}.
In the analysis of these diagrams the input  supersymmetry parameters consist of the following
\begin{gather}
M_{\tilde e L}, M_{\tilde \nu_e}, M_{\tilde e},  \mu, \tan\beta,  M_1, M_2,  A_e, A_{\nu_e}
\end{gather}
where $M_{\tilde e L}$ etc are the soft scalar masses, $M_1, M_2$ are the gaugino masses in the
$U(1)$ and $SU(2)$ sectors, $A_e$ etc are the trilinear couplings, $\mu$ is the Higgs mixing parameter
which enters the superpotential as $\mu H_1 H_2$,  where $H_2$ gives mass to the up quarks and $H_1$ gives
mass to the down quarks and the leptons, while $\tan\beta$ is the ratio of the Higgs VEVs so that
$\tan\beta= <H_2>/<H_1>$ (see Appendix A for discussion of the soft parameters). 
Further,  $\mu$, $M_1$, $M_2$, and the trilinear coupling $A_k$ 
 are complex and we define their phase so that 
\begin{gather}
\mu= |\mu| e^{i\alpha_\mu}, ~~M_i= |M_i| e^{\alpha_i}, i=1,2\\
A_k= |A_k| e^{i\alpha_{A_k}}, ~~ k=e, \nu_e \  .
\end{gather}
The analysis of the diagrams of \cref{fig1} involves  electron-chargino-sneutrino interactions and the electron-neutralino-slepton  interactions. For the chargino-sneutrino exchange diagrams one has 

\beqn
 d_e^{\chi^{-}}= \frac{\alpha_{em}}{ 4\pi \sin^2\theta_W} \frac{k_e}{m_{\tilde \nu_e}^2} 
 \sum _{i=1}^2 m_{\tilde \chi^-_i} Im (U_{i2}^* V_{i1}^*) F \left(\frac{  m^2_{\tilde \chi^-_i}}{ m_{\tilde \nu_e}^2}  \right)
\label{2.1}
\eeqn
where $F(x)$ is a form factor defined by 
\begin{equation}
F(x)=  \frac{1}{2(1-x)^2} \left(3- x + \frac{2 \ln x}{1-x}\right)
\label{2.2}
\end{equation}
and 
\begin{gather}
\kappa_e = \frac{m_e}{ \sqrt 2 m_W \cos\beta}.
\label{2.3}
\end{gather}
Here $U,V$ diagonalize the chargino mass matrix $M_C$ so that 

\begin{equation}
U^* M_C V= {\rm diag} (m_{\tilde \chi_1^-}, m_{\tilde \chi_2^-}).
\label{2.4}
\end{equation}
For the neutralino-slepton exchange diagrams one finds 
\beqn
d_e^{\tilde \chi^0}= \frac{\alpha_{em}}{ 4\pi \sin^2\theta_W} 
\sum_{k=1}^2 \sum_{i=1}^4 Im(\eta_{eik})  \frac{m_{\tilde \chi_i^0}}{M_{\tilde fk^2}} Q_{\tilde f} 
G \left(\frac{  m^2_{\tilde \chi^-_i}}{ m_{\tilde \nu_e}^2}  \right)
\label{2.5}
\eeqn
where $G(x)$ is a form factor defined by 
\begin{equation}
G(x) =  \frac{1}{2(1-x)^2} \left(1+ x + \frac{2 x \ln x}{1-x}\right)
\label{2.6}
\end{equation}
where 
\begin{gather}
\eta_{eik} = \left[- \sqrt 2 
\left\{ \tan\theta_W (Q_e- T_{3e}) X_{1i} + T_{3i} X_{2i} \right\} D_{e1k}^* + \kappa_e 
X_{bi} D_{e2k}^*\right]\\
\left (\sqrt 2 \tan\theta_W Q_e X_{1i} D_{e2k} - \kappa_e X_{bi} D_{e1k}\right).
\label{2.7}
\end{gather}
 where $b=3$ and $T_{3e}= -1/2$. Further, $X_{ij}$ are elements of the matrix $X$ which diagonalizes the
 neutralino mass matrix $M_{\chi^0}$  so that
 \begin{equation}
X^T M_{\chi^0} X= {\rm diag} \left( m_{\tilde \chi_1^0},  m_{\tilde \chi_2^0}, m_{\tilde \chi_3^0}, m_{\tilde \chi_4^0}\right)\ ,
\label{2.8}
\end{equation}
and $D_e$ diagonalizes the scalar electron mass $^2$ matrix so that 
\begin{gather}
\tilde e_L= D_{e11} \tilde e_1 + D_{e12} \tilde e_2, 
~\tilde e_R= D_{e21} \tilde e_1 + D_{e22} \tilde e_2
\label{2.9}
\end{gather}
where $\tilde e_1$ and $\tilde e_2$ are the selectron  mass eigenstates. 
 In \cref{fig2} we give a numerical analysis of  the electron EDM as a function of $m_0$.  
 Here one finds that the current constraint on the electron EDM allows one to probe the $m_0$ region in the
 tens of TeV while improvement in the sensitivity by a factor of 10 or more will allow one to extend the
 range up to 100 TeV - 1 PeV.
 
  \vspace{0.5cm}
 
\section{EDM Analysis by inclusion of a vector generation in MSSM\label{sec3}}
Next we discuss the case when we include a vectorlike leptonic multiplet which mixes with the three
generations of leptons. In this case the mass eigenstates will be linear combinations of the three
generations plus the vector like generation which includes mirror particles. The details of
the model and its interactions are given in {Appendices} A-C. Here we discuss the contribution of the 
model to the electron EDM. These contributions arise from four sources:  the chargino exchange, the neutralino exchange,
the  $W$ boson exchange and the $Z$ boson exchange.

\begin{figure}[t]
\begin{center}
{\rotatebox{0}{\resizebox*{14cm}{!}{\includegraphics{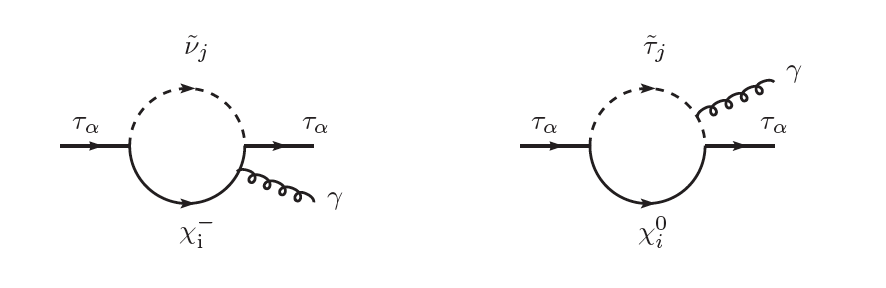}}\hglue5mm}}\\
{\rotatebox{0}{\resizebox*{14cm}{!}{\includegraphics{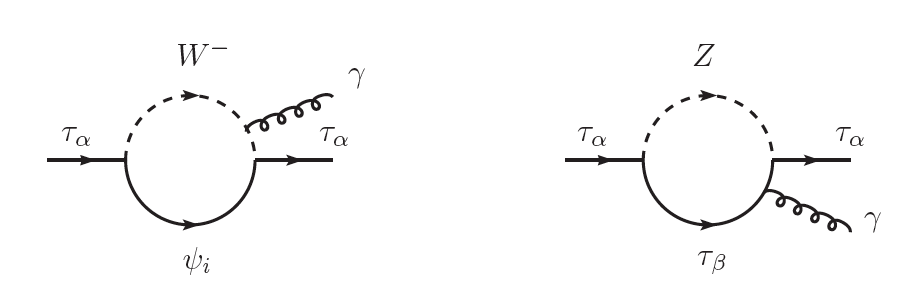}}\hglue5mm}}
\caption{
Upper diagrams: Supersymmetric contributions to the leptonic EDMs
arising from the exchange of the charginos, sneutrinos
and mirror sneutrinos (upper left) and the exchange of neutralinos, sleptons, and mirror sleptons (upper right) 
inside the loop. Lower diagrams: 
Non-supersymmetric  diagrams that contribute to the leptonic EDMs
via the exchange of  the $W$, the sequential and vector like neutrinos (lower left) and the exchange of the 
$Z$, the sequential and vector like charged leptons (lower right).}
 \label{fig3}
\end{center}
\end{figure}
Using the interactions given in Appendix B  the chargino contribution  is given by

\begin{align}
d_{\alpha}^{\chi^{+}}&=-\frac{1}{16\pi^2}\sum_{i=1}^{2}\sum_{j=1}^{8}\frac{m_{\chi^{+}_i}}{m^2_{\tilde\nu_{j}}}\text{Im}(C^{L}_{\alpha ij}C^{R*}_{\alpha ij})
F\left(\frac{m^{2}_{{\chi^{+}}_{i}}}{m^{2}_{\tilde\nu_{i}}}\right) 
\label{3.1}
\end{align}
where the functions $C^L$ and $C^R$ are given in   Appendix B and 
 the form factor $F(x)$  is given by   { \cref{2.2}}. 
Using the interactions given in  Appendix B the neutralino contribution  is given by

\begin{align}
d_{\alpha}^{\chi^{0}}&=-\frac{1}{16\pi^2}\sum_{i=1}^{4}\sum_{j=1}^{8}\frac{m_{\chi^{0}_i}}{m^2_{\tilde\tau_{j}}}\text{Im}(C'^{L}_{\alpha ij}C'^{R*}_{\alpha ij})
G\left(\frac{m^{2}_{{\chi^{0}}_{i}}}{m^{2}_{\tilde\tau_{i}}}\right) 
\end{align}
where the functions $C^{'L}$ and $C^{'R}$ are defined in  Appendix B and the
 form factor $G(x)$  is given by Eq. (\ref{2.6}).
 The contributions to the lepton electric moment from the $W$ and $Z$ exchange arise from similar loops.  Using the interactions given in  Appendix B the contribution arising from
 the $W$ exchange diagram is given by

\begin{align}
d_{\alpha}^{W}&=\frac{1}{16\pi^2}\sum_{i=1}^{4}\frac{m_{\psi^{+}_i}}{m^2_W}\text{Im}(C^{W}_{Li\alpha }C^{W*}_{R i\alpha })
I_1\left(\frac{m^{2}_{{\psi}_{i}}}{m^{2}_{W}}\right) 
\end{align}
where the functions $C_L^W$ and $C_R^W$ are given in  Appendix B and the 
 form factor $I_1$  is given by

\begin{align}
I_1(x)&=\frac{2}{(1-x)^{2}}\left[1-\frac{11}{4}x +\frac{1}{4}x^2-\frac{3 x^2\ln x}{2(1-x)} \right]
\end{align}

The $Z$ boson exchange diagram contribution is given by
\begin{align}
d_{\alpha}^{Z}&=-\frac{1}{16\pi^2}\sum_{\beta=1}^{4}\frac{m_{\tau_\beta}}{m^2_Z}\text{Im}(C^{Z}_{L\alpha\beta }C^{Z*}_{R \alpha\beta })
I_2\left(\frac{m^{2}_{\tau_{\beta}}}{m^{2}_{Z}}\right) 
\end{align}
where the functions $C_L^Z$ and $C_R^Z$ are defined in  Appendix B and  
where the form factor $I_2$  is given by

\begin{align}
I_2(x)&=\frac{2}{(1-x)^{2}}\left[1+\frac{1}{4}x +\frac{1}{4}x^2+\frac{3 x\ln x}{2(1-x)} \right]
\label{23}
\end{align}

\section{Numerical analysis and results\label{sec4}}
\begin{figure}[H]
\begin{center}
{\rotatebox{0}{\resizebox*{7.5cm}{!}{\includegraphics{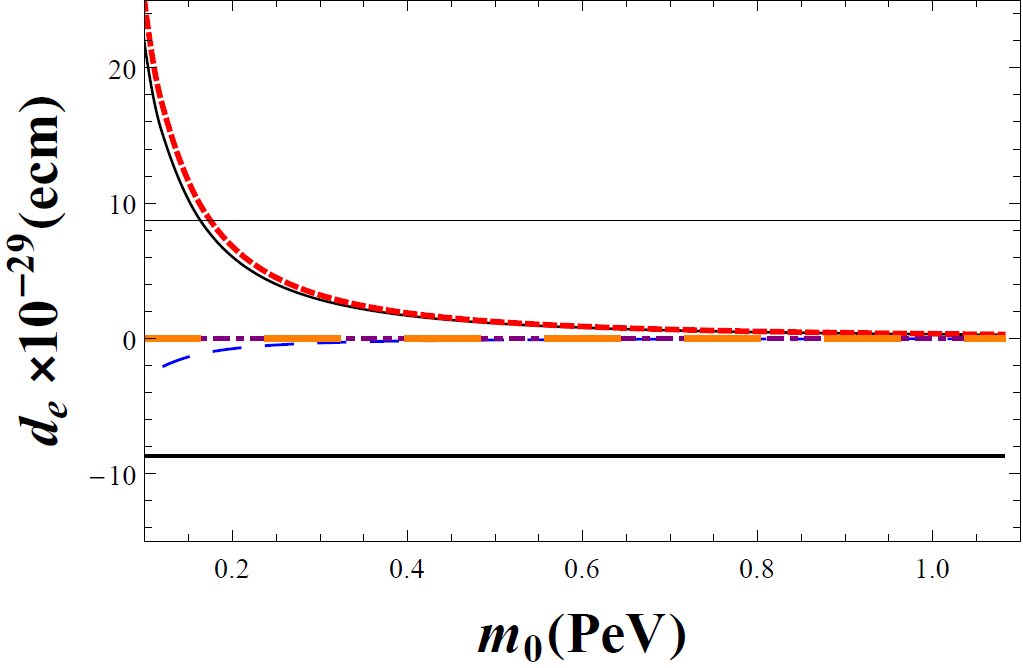}}\hglue5mm}}
{\rotatebox{0}{\resizebox*{7.5cm}{!}{\includegraphics{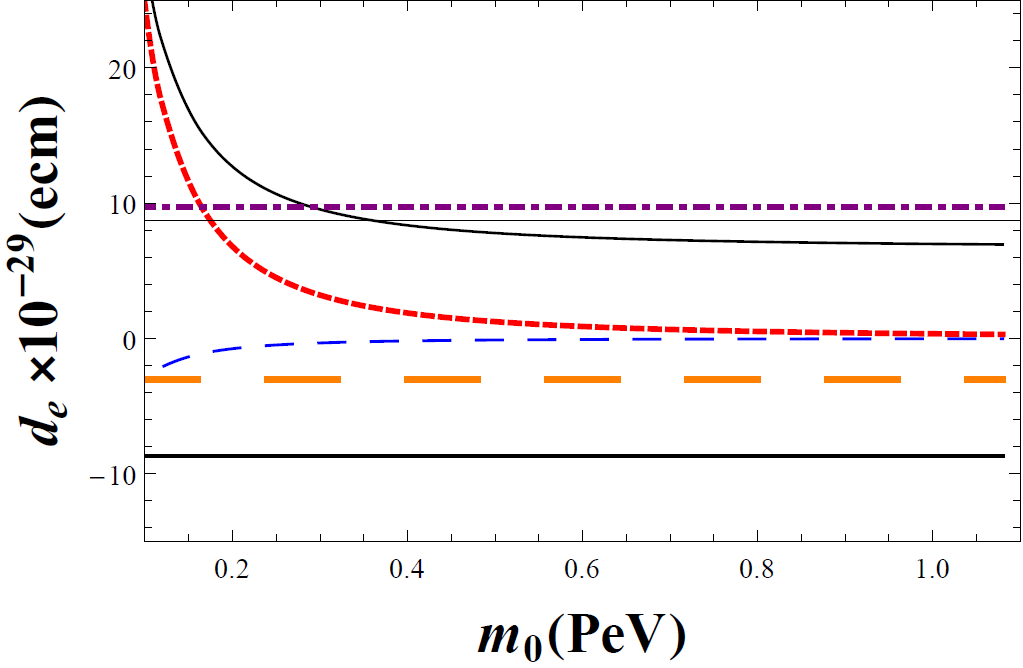}}\hglue5mm}}\\
\caption{
Left Panel:   Exhibition of the individual contributions to the EDM of the electron when there is no 
mixing between the vectorlike generation and the three regular generations. The parameters chosen
for this case are the same as for the solid curve in \cref{fig2} where $\alpha_\mu=-0.5$.
As expected the
contributions from the W-exchange (the long-dashed curve in orange) and the Z -exchange (dot-dashed purple curve)
 give vanishing contribution in this case, and the entire contribution arises from the 
chargino-exchange (the small-dashed curve in red) and the neutralino -exchange( the medium-dashed blue curve).
Right Panel: The parameter point chosen is the same as for the left panel except that mixing of the 
vectorlike generation with the regular three generations is allowed. The additional parameters chosen
are  $m_N=250, m_E=380$ and the f couplings set to $\left|f_3\text{$|$ = }\right.7.20\times 10^{-6}\text{ ,  $|$}f_3'\text{$|$ = }1.19\times 10^{-4}\text{ ,  $|$}f_3''\text{$|$ = }1.55\times 10^{-5}\text{ ,  $|$}f_4\text{$|$ = }8.13\times 10^{-4}\text{ ,  $|$}f_4'\text{$|$ = }3.50\times 10^{-1}\text{ ,  $|$}f_4''\text{$|$ = }6.29\times 10^{-1}\text{ ,  $|$}f_5\text{$|$ = }8.82\times 10^{-5}\text{ ,  $|$}f_5'\text{$|$ = }5.36\times 10^{-5}\text{ ,  $|$}f_5''\text{$|$ = }1.27\times 10^{-5}$. Their corresponding CP phases set to $\chi _3\text{ = }9.71\times 10^{-1}\text{ ,  }\chi _3'\text{ = }7.86\times 10^{-1}\text{ ,  } \chi _3''\text{ = }7.89\times 10^{-1}\text{ ,  }\chi _4\text{ = }7.66\times 10^{-1}\text{ ,  }\chi _4'\text{ = }8.38\times 10^{-1}\text{ ,  }\chi _4''\text{ = }8.23\times 10^{-1}\text{ ,  }\chi _5\text{ = }7.70\times 10^{-1}\text{ ,  }\chi _5'\text{ = }1.47\text{ ,  }\chi _5''\text{ = }7.82\times 10^{-1}$. All masses are in GeV,  phases in rad and EDM in $e$cm. }
\label{fig4}
\end{center}
\end{figure}

{
We discuss now the numerical analysis for the EDM of the electron in the model given in Section 3.
The parameter space of the model of Section 3  is rather large. In addition to the MSSM parameters, one has 
the parameters arising from the vectorlike multiplet and its mixings with the  standard model generations
of quarks and leptons.
Thus as in  MSSM  here also we look at slices of the parameter space
to show that interesting new physics exists in these regions.}
Thus for simplicity in the analysis we assume $A_{\nu_{\tau}}= A_{\nu_{\mu}}= A_{\nu_{e}}=A_{N}=A_{0}^{\tilde{\nu}}$ and
   $m_{0}^{\tilde{\nu}^{2}}={M}_{\tilde N}^{2}={M}_{\tilde\nu_{\tau}}^{2}={M}_{\tilde\nu_{\mu}}^{2}={M}_{\tilde\nu_{e}}^{2}$ 
   in the sneutrino mass squared matrix (see \cref{13}).  We also assume 
  $m_{0}^{2}={M_{\tilde\tau L}}^{2}={M}_{\tilde E}^{2}={M}_{\tilde \tau}^{2}={M}_{\tilde \chi}^{2}={M}_{\tilde \mu L}^{2}={M}_{\tilde\mu}^{2}={M}_{\tilde e L}^{2}={M}_{\tilde e}^{2}$ and $A_{0}=A_{\tau}=A_{E}=A_{\mu}=A_{e}$ in the slepton mass squared matrix (see \cref{13}).  
  {The assumed  masses for the new leptons are consistent with the lower limits given by the Particle Data Group\cite{pdg}.}
    In \cref{fig2} we investigated $d_e$ in MSSM as a function of $m_0$ when there were no
  mixing of the ordinary leptonic generations with the vectorlike generation. We wish now to switch
  on a small mixing with the vector like generation and see what effect it has on the electron EDM.
  To this end we focus on one curve in \cref{fig2} which we take to be the solid curve (the case
  $\alpha_\mu=-0.5$). For this case we plot the individual contributions to $d_e$ in the left panel 
  of \cref{fig4}. Here one finds that the largest contribution to $d_e$ arises from the chargino
  exchange while the 
   neutralino exchange produces a much smaller contribution 
  and as expected the $W$ and $Z$ exchanges do not contribute. \\
  
  Next we turn on
  a small coupling between the vector like generation and the three generations of leptons.
    The analysis for this case is given in the right panel of \cref{fig4}.
   The turning on of the mixings has the following effect: the supersymmetric contribution is modified only modestly
  and its general feature remains as in the left panel. However, now because of mixing 
  with the vectorlike generation the contribution from the $W$ and $Z$ exchange is  non-vanishing 
  and in fact is very significant. Further, unlike the chargino and the neutralino exchange
  contribution the $W$ and $Z$ exchange contribution does not depend on $m_0$ 
  as exhibited in  \cref{fig4}. Thus as $m_0$ gets large the 
  supersymmetric contributions becomes much smaller than that of the $W$ and $Z$ exchange 
  contribution. For this reason, $d_e$ is dominated by the $W$ and  $Z$ exchange.
  This phenomenon is exhibited  in further detail in  \cref{table1} which is 
  done for the same  set of parameters as the right panel of \cref{fig4} except that  $m_0=1.1$ PeV.  
 Here column (i) gives the individual contributions for the case (i) of no mixing where 
 $W$ and $Z$ contributions vanish, and the non-vanishing contributions arise from chargino
 and neutralino exchange. 
 Column (ii) exhibits the individual contributions when the mixings with the 
 vector like generation are turned on. Here one finds that the supersymmetric contributions
 from the chargino and neutralino exchanges are essentially unchanged 
 from the case of no mixing  but the 
 contributions from the $W$ and $Z$ exchanges are now non-zero and are in fact much larger
 than the chargino and neutralino exchange contributions.  The reason for the non-vanishing
 contribution from the $W$ and $Z$ exchanges is due to
the mixings with vector like generation whose couplings are complex and carry CP violating phases.

\begin{table}[H] \centering
\begin{tabular}{ccc}
\toprule\toprule
 & (i) Case of no mixing & (ii) Case of mixing \\ \cmidrule{2-3}
$d_e^{\chi^+}$ & $2.82\times 10^{-30}$ & $2.82 \times 10^{-30}$  \\
$d_e^{\chi^0}$ & $-2.53 \times 10^{-31}$ & $-2.53\times10^{-31}$  \\
$d_e^W$ & $0$ & $9.72 \times 10^{-29}$  \\
$d_e^Z$ & $ 0$ & $-3.05 \times 10^{-29}$  \\ \hline
 $d_e$ & $2.57\times 10^{-30}$ & $6.93 \times 10^{-29}$  \\
\bottomrule \bottomrule
\end{tabular}
\caption{Column (i):
An exhibition of the individual contributions to $d_e$ arising from the chargino, neutralino, W and Z boson exchanges and their sum $d_e$  for the case when there is no mixing among the generations.
The parameters chosen are the same as for the solid curve   ($\alpha_\mu=-0.5$ rad) of \cref{fig2}
 where  $m_0$ is set to 1.1 PeV. 
 Column (ii): 
 The analysis of column (ii) has the same set of parameters as the left panel except that inter-generational couplings are  allowed. Here the   couplings  $f_3, f_3'$, $f_3'', f_4, f_4'$, $f_4'', f_5, f_5'$, and $f_5''$ are the same 
  as the ones in the right panel of \cref{fig4}. The fermion masses for the vectorlike generation 
  are $m_N=250$ and $m_E=380$ GeV. The  EDM is in $e$cm units.} 
\label{table1}
\end{table}
  
\begin{figure}[h]
\begin{center}
{\rotatebox{0}{\resizebox*{10cm}{!}{\includegraphics{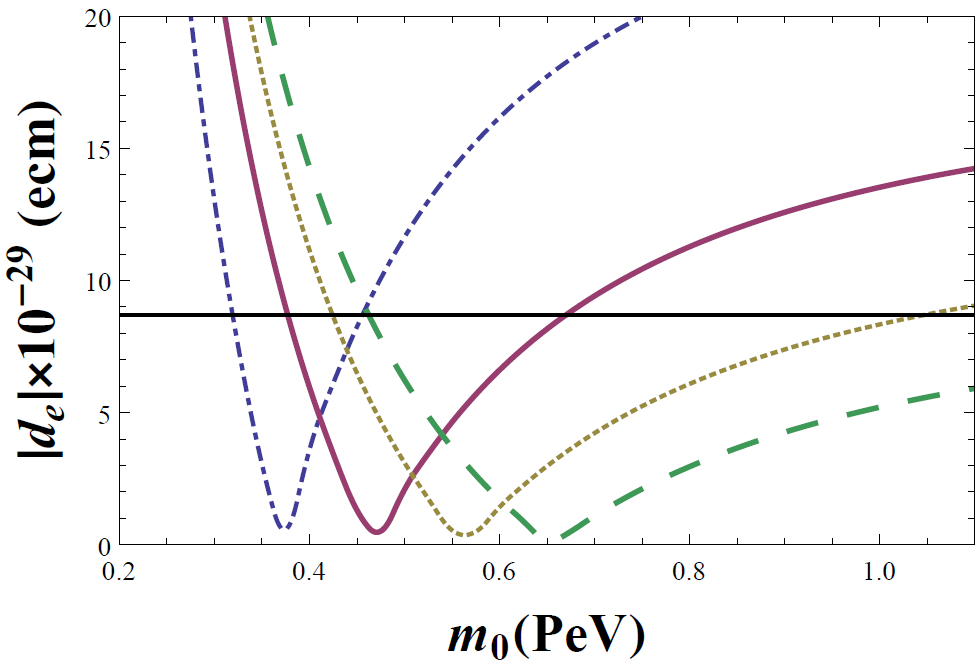}}\hglue5mm}}
\caption{An exhibition of the dependence of $|d_e|$ on $m_0$ for various vectorlike masses. The curves correspond to $m_N=m_E=150$ (dotdashed), $m_N=m_E=200$ (solid), $m_N=m_E=250$ (dotted), $m_N=m_E=300$ (dashed). The parameters are $\text{$|\mu |$ = }4.1\times 10^2\text{ ,  $|$}M_1\text{$|$ = }2.8\times 10^2\text{ ,  $|$}M_2\text{$|$ = }3.4\times 10^2\text{ , $|$}A_0\text{$|$ = }3\times 10^6\text{ ,  }m_0^{\tilde{\nu }}\text{ = }4\times 10^6\text{ ,  $|$}A_0^{\tilde{\nu }}\text{$|$ = }5\times 10^6\text{ ,  tan$\beta $ = }50$. The CP phases are $\theta_{\mu }\text{ = }1\text{ ,  }\alpha _1\text{ = }1.26\text{ ,  }\alpha _2\text{ = }0.94\text{ ,  }\alpha _{A_0}\text{ = }0.94\text{ ,  }\alpha _{A_0^{\tilde{\nu }}}\text{ = }1.88$. The f couplings are $\left|f_3\text{$|$ = }\right.3.01\times 10^{-5}\text{ ,  $|$}f_3'\text{$|$ = }8.07\times 10^{-6}\text{ ,  $|$}f_3''\text{$|$ = }2.06\times 10^{-5}\text{ ,  $|$}f_4\text{$|$ = }8.13\times 10^{-4}\text{ ,  $|$}f_4'\text{$|$ = }3.50\times 10^{-1}\text{ ,  $|$}f_4''\text{$|$ = }6.29\times 10^{-1}\text{ ,  $|$}f_5\text{$|$ = }6.38\times 10^{-5}\text{ ,  $|$}f_5'\text{$|$ = }1.03\times 10^{-6}\text{ ,  $|$}f_5''\text{$|$ = }2.44\times 10^{-8}$. Their corresponding CP phases are $\chi _3\text{ = }7.91\times 10^{-1}\text{ ,  }\chi _3'\text{ = }7.87\times 10^{-1}\text{ ,  } \chi _3''\text{ = }7.78\times 10^{-1}\text{ ,  }\chi _4\text{ = }7.66\times 10^{-1}\text{ ,  }\chi _4'\text{ = }8.38\times 10^{-1}\text{ ,  }\chi _4''\text{= }8.23\times 10^{-1}\text{ ,  }\chi _5\text{ = }7.57\times 10^{-1}\text{ ,  }\chi _5'\text{= }7.54\times 10^{-1}\text{ ,  }\chi _5''\text{= }7.83\times 10^{-1}$. All masses are in GeV,  phases in rad, and $d_e$ in $e$cm. }
\label{fig5}
\end{center}
\end{figure}

In \cref{fig5} we give an analysis of the electron EDM as a function of $m_0$ for different pairs of
fermion masses for the vectorlike generation. The fermion masses for the vectorlike generation
lies in the range 150-300 GeV. Here we find that $d_e$ is very sensitive to the fermion masses for the
vector like generation. The dependence of $|d_e|$ on $m_0$ shows a turn around where $|d_e|$ first
decreases and then increases. This is easily understood as follows: As discussed already for the 
case of \cref{fig4} the supersymmetric contribution is very sensitive to $m_0$ since the sleptons 
that enter in the supersymmetric diagrams get large as $m_0$ gets large and consequently 
the SUSY contributions become negligible as $m_0$ gets large. However, also as already 
discussed the $W$ and $Z$ exchange contributions are not affected by $m_0$. Thus at low
values of $m_0$, the supersymmetric contribution is large and of opposite sign to the $W$ and $Z$ exchange 
 contribution
in this region of the parameter space
 which leads to a cancellation between the two thus a falling behavior 
of $|d_e|$. However, as $m_0$ increases the SUSY contribution dies out and the $W$ and $Z$ contribution
take over which explains the turn around. This turn around is exhibited for two values of $m_0$
around the minimum
in \cref{table2}. Here we consider the parameter point $m_N=m_E=200$  GeV in \cref{fig4} 
for the sample points $m_0=0.4$ PeV and $m_0=0.6$ PeV. Comparison of columns (i) and (ii) in 
\cref{table2} 
shows that the chargino and the neutralino exchange contribution vary  in a significant way 
while the $W$ and $Z$ exchange contribution  is  unchanged. Consequently 
$d_e=-5.96\times 10^{-29}$ $e$cm for column (i) and $d_e= 6.61\times 10^{-29}$ $e$cm
for column (ii). Thus we see that the $d_e$ has switched the sign in going from
$m_0=0.4$ PeV to $m_0=0.6$ PeV which means that $d_e$ has gone through a zero
which explains the turn around of $|d_e|$ in \cref{fig5}.\\

\begin{figure}[H]
\begin{center}
{\rotatebox{0}{\resizebox*{7.5cm}{!}{\includegraphics{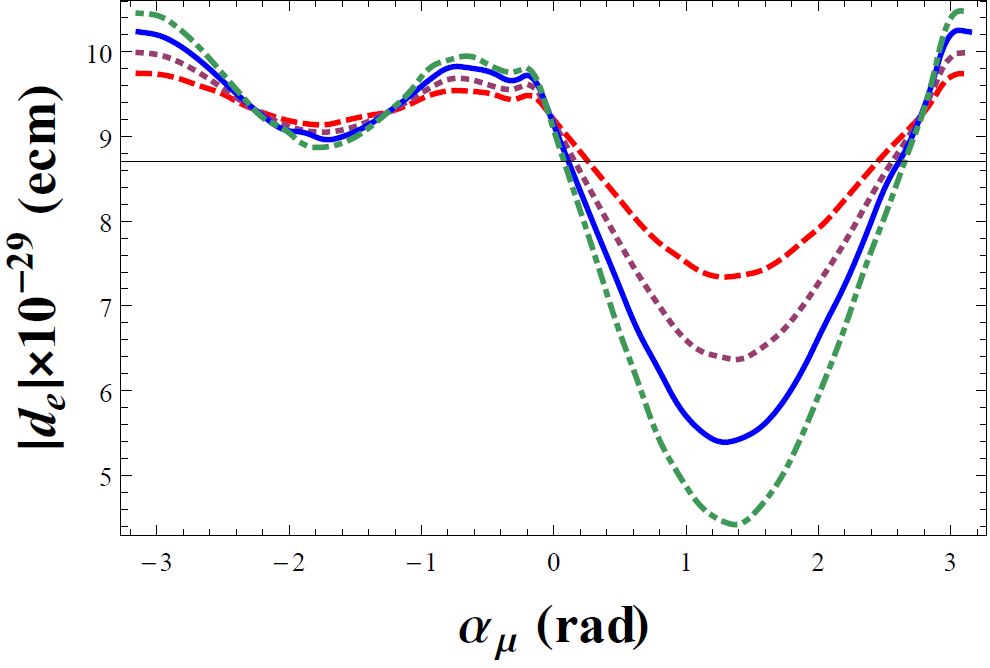}}\hglue5mm}}
\caption{An exhibition of the dependence of $|d_e|$ on $\alpha_\mu$ for various $\tan\beta$. The curves correspond to $\tan\beta=20$ (dashed), $\tan\beta=30$ (dotted), $\tan\beta=40$ (solid), and $\tan\beta=50$ (dotdashed).  The parameters used are $\text{$|\mu |$ = }3.9\times 10^2$ $\text{ ,  $|$}M_1\text{$|$ = }3.1\times 10^2\text{ ,  $|$}M_2\text{$|$ = }3.6\times 10^2\text{ ,  }m_N\text{ = }340\text{ ,  }m_E\text{ = }250\text{ ,  }m_0\text{ = }1.1\times 10^6\text{ ,  $|$}A_0\text{$|$ = }3.2\times 10^6\text{ ,  }m_0^{\tilde{\nu }}\text{ = }4.3\times 10^6$ $\text{ ,  $|$}A_0^{\tilde{\nu }}\text{$|$ =} 5.1\times 10^6$  $ \text{ , }\alpha _1\text{ = }1.88 $ $ \text{ ,  }\alpha _2\text{ = }1.26 
\text{ ,  }\alpha _{A_0}\text{ = }0.94\text{ ,  }\alpha _{A_0^{\tilde{\nu }}}\text{ = }1.88$. $\text{The mixings are $|$}f_3\text{$|$ = }2.88\times 10^{-4}\text{ ,  $|$}f_3'\text{$|$ = }8.19\times 10^{-6}\text{ ,  $|$}f_3''\text{$|$ = }9.19\times 10^{-5}\text{ ,  $|$}f_4\text{$|$ = }8.13\times 10^{-4}\text{ ,  $|$}f_4'\text{$|$ = }3.50\times 10^{-1}\text{ ,  $|$}f_4''\text{$|$ = }1.29\times 10^{-1}\text{ ,  $|$}f_5\text{$|$ = }5.75\times 10^{-6}\text{ ,  $|$}f_5'\text{$|$ = }1.00\times 10^{-5}\text{ ,  $|$}f_5''\text{$|$ = }2.49\times 10^{-7}\text{ ,  }\chi _3\text{ = }7.74\times 10^{-1}\text{ ,  }\chi _3'\text{ = }7.73\times 10^{-1}\text{ ,  } \chi _3''\text{ = }7.86\times 10^{-1}\text{ ,  }\chi _4\text{ = }7.6\times 10^{-1}\text{ ,  }\chi _4'\text{ = }8.40\times 10^{-1}\text{ ,  }\chi _4''\text{ = }8.20\times 10^{-1}\text{ ,  }\chi _5\text{ = }7.51\times 10^{-1}\text{ ,  }\chi _5'\text{ = }8.19\times 10^{-1}\text{ ,  }\chi _5''\text{= }8.03\times 10^{-1}$. All masses are in GeV,  phases in rad, and $d_e$ in $e$cm. }
\label{fig6}
\end{center}
\end{figure}

\begin{figure}[h]
\begin{center}
{\rotatebox{0}{\resizebox*{7.5cm}{!}{\includegraphics{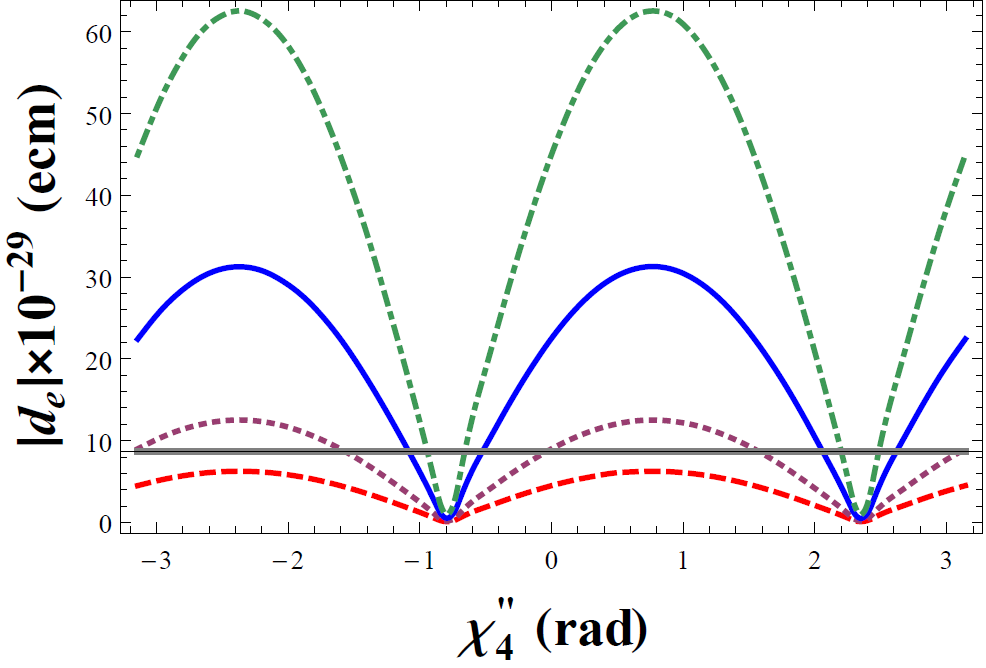}}\hglue5mm}}
\caption{An exhibition of the dependence of $|d_e|$ on $\chi_4^{''}$ for various $f_4^{''}$. The curves correspond to $f_4^{''}$ of $0.1$ (dashed),  $0.2$ (dotted), $0.5$ (solid), $1$ (dotdashed). The other parameters  are $\text{$|\mu |$ = }1.1\times 10^6\text{ ,  $|$}M_1\text{$|$ = }2.8\times 10^6\text{ ,  $|$}M_2\text{$|$ = }3.4\times 10^6\text{ ,  }m_N\text{ = }250\text{ ,  }m_E\text{ = }380\text{ ,  }m_0\text{ = }1.1\times 10^6\text{ ,  $|$}A_0\text{$|$ = }3.2\times 10^6\text{ ,  }m_0^{\tilde{\nu }}\text{ = }1.4\times 10^6\text{ ,  $|$}A_0^{\tilde{\nu }}\text{$|$ = }5.1\times 10^6\text{ , }\alpha _1\text{ = }1.26\text{ ,  }\alpha _2\text{ = }0.94\text{ ,  }\alpha _{A_0}\text{ = }0.94\text{ ,  }\alpha _{A_0^{\tilde{\nu }}}\text{ = }1.88\text{ , tan$\beta $ = }30$. $\text{The mixings are $|$}f_3\text{$|$ = }2.93\times 10^{-4}\text{ ,  $|$}f_3'\text{$|$ = }8.19\times 10^{-6}\text{ ,  $|$}f_3''\text{$|$ = }9.15\times 10^{-5}\text{ ,  $|$}f_4\text{$|$ = }8.13\times 10^{-1}\text{ ,  $|$}f_4'\text{$|$ = }3.50\times 10^{-1}\text{ , $|$}f_5\text{$|$ = }5.08\times 10^{-6}\text{ ,  $|$}f_5'\text{$|$ = }9.98\times 10^{-6}\text{ ,  $|$}f_5''\text{$|$ = }2.56\times 10^{-7}\text{ ,  }\chi _3\text{ = }7.86\times 10^{-1}\text{ ,  }\chi _3'\text{ = }7.80\times 10^{-1}\text{ ,  } \chi _3''\text{ = }8.02\times 10^{-1}\text{ ,  }\chi _4\text{ = }7.6\times 10^{-1}\text{ ,  }\chi _4'\text{ = }8.4\times 10^{-1}\text{ , }\chi _5\text{ = }7.39\times 10^{-1}\text{ ,  }\chi _5'\text{ = }7.82\times 10^{-1}\text{ ,  }\chi _5''\text{ = }7.82\times 10^{-1}$. All masses are in GeV,  phases in rad and $d_e$ in $e$cm.
}
\label{fig7}
\end{center}
\end{figure}

In \cref{fig6} we exhibit the dependence of $|d_e|$ on the phase $\alpha_\mu$ which is the phase of the 
Higgs mixing parameter $\mu$. The dependence of   $|d_e|$ on $\alpha_\mu$ arises from various sources. Thus the 
slepton masses as well as the chargino and the neutrino masses that enter in the supersymmetric
loop contribution have a dependence on $\alpha_\mu$ which makes a simple explanation of the
dependence on this parameter less transparent. A numerical analysis exhibiting the dependence of
$|d_e|$ on $\alpha_\mu$ is given in \cref{fig6}. The analysis is done for different $\tan\beta$ ranging from
$\tan\beta=20$ to $\tan\beta=50$.  A similar analysis of the dependence of $|d_e|$ on $\chi_4''$ for 
various values of $f_4''$ is given in \cref{fig7}. The sharp dependence of $|d_e|$ on $\chi_4''$ is not difficult
to understand. Unlike the case of  the dependence of $|d_e|$ 
on $\alpha_\mu$ which arises mainly from the supersymmetric sector, here the dependence of $|d_e|$ on 
$\chi_4''$ arises from the non-supersymmetric sector via the exchange of $W$ and $Z$ bosons.
The {SUSY} contribution dependence is limited by the smallness of $| f_4''|$ compared to the other masses in the slepton mass$^2$ matrix.
 The non-supersymmetric
contribution is directly governed by $f_3'', f_4'', f_5''$ as can be seen from Eq.(\ref{7aa}) and Eq.(\ref{7bb}).
Here setting $f_3''=f_4''=f_5''=0$ puts the mass matrices in a block diagonal form where the first generation
totally decouples from the vector like generation. This clearly indicates that the effect of variation in
$|f_3''|, |f_4''|, |f_5''|$ and their phases, $\chi_3'', \chi_4'', \chi_5''$ will be strong.
This is what the analysis of  \cref{fig7} indicates. Aside from the variations of the $W$ and $Z$ contributions
on $\chi_4''$, there is also  a constructive/destructive interference between the $W$ and the $Z$ contributions 
as $\chi_4''$ varies which explains the rapid variations of $|d_e|$ with $\chi_4''$ in \cref{fig7}.\\

Finally, the effect of mixing of the vectorlike generation with the three lepton generations has negligible effect
on the standard model predictions in the leptonic sector at the tree level. However, it does
affect the neutrino sector. Specifically taking the mixings into account the analysis presented here
satisfies the  constraint on the sum of the neutrino masses arising from the Planck Satellite experiment~\cite{Schwetz:2008er}  so that 
 \beq
\sum_{i=1}^3m_{\nu_{i}}<0.85 \ {\rm eV} \ , 
\label{6.1a}
\eeq
where we assume $\nu_i$ (i=1,2,3) to be the mass eigenstates with eigenvalues $m_{\nu_i}$.
Further,  the neutrino oscillations constraint on the neutrino mass squared differences~\cite{Schwetz:2008er}
are also satisfied, i.e., the constraints
\begin{gather}
\label{6.1b}
\Delta m^2_{31}\equiv m_3^2-m_1^2= 2.4^{+0.12}_{-0.11} \times 10^{-3} ~{\rm eV}^2  \ , \\
\Delta m_{21}^2\equiv m_2^2- m_1^2= 7.65^{+0.23}_{-0.20} \times 10^{-5}~{\rm eV}^2. 
\label{6.1c}
\end{gather}

The analysis given in this section respect all of the collider, i.e., LEP and LHC, constraints. 
Specifically the lower limits on heavy lepton masses is around 100 GeV\cite{pdg} and masses of 
$m_E$ and $m_N$ used here respect these limits.  However,
in addition there are flavor constraints to consider. Here the constraint $\mu\to e+\gamma$ is the most 
stringent constraint.  Thus 
the above framework allows the process $\mu\to e+\gamma$ for which the current upper 
limit from experiment is \cite{pdg} $4.4\times 10^{-12}$. The analysis of this process requires 
the mixing of the vectorlike generation with all the three generations. A similar analysis but for the
$\tau\to \mu + \gamma$ was given in \cite{Ibrahim:2012ds} and it was found that the model with a vector like 
generation can produce a branching ratio for this process which lies below the current experimental 
limit for that process but could be accessible in improved experiment .  In that analysis the scalar masses were in the sub TeV region. However, in the present case we are
interested in the PeV size scalar masses. 
From Figure 3 of \cite{Ibrahim:2012ds}, we see that for heavy scalars, the branching ratio decreases rapidly as the masses increase and since we are interested in the 
PeV size scalars we expect that the $\mu\to e + \gamma$ experimental upper limits would be easily 
satisfied. A full treatment of the processes is, however, outside the scope of this work and will be 
discussed elsewhere.

\begin{table} \centering
\begin{tabular}{ccc}
\toprule\toprule
 & (i) $m_{0}= 0.4$ PeV  & (ii) $m_{0}= 0.6$ PeV   \\ \cmidrule{2-3}
$d_e^{\chi+}$ & $-2.38\times 10^{-28}$ & $-1.13 \times 10^{-28}$  \\
$d_e^{\chi0}$ & $-9.18\times 10^{-31}$ & $-4.08\times 10^{-31}$   \\
 $d_e^W$ & $2.72\times 10^{-28}$ & $2.72 \times 10^{-28}$  \\
$d_e^Z$ & $-9.31 \times 10^{-29}$ & $-9.31 \times 10^{-29}$   \\ \hline
$d_e$ & $-5.96 \times 10^{-29}$ & $6.61\times 10^{-29}$  \\
\bottomrule \bottomrule
\end{tabular}
\caption{An exhibition of the individual  contributions to the electric dipole moment of the electron arising from the chargino exchange, neutralino exchange, W boson exchange and Z boson exchange. 
The last row gives the total EDM $d_e$ where
$d_e= d_e^{\chi+} + d_e^{\chi0}  +  d_e^{W} + d_e^{Z}$.
The analysis is for the solid curve of \cref{fig5} where $m_N=m_E=200$ when (i) $m_0=0.4$ PeV, (ii) $m_0=0.6$ PeV.
The EDM is in $e$cm units.}
\label{table2}
\end{table}

\section{Conclusion\label{sec5}}
In the future the exploration of high scale physics on the energy frontier  
 will be limited by the  capability on the highest  energy 
that accelerators can achieve.  Thus the upgraded LHC will achieve an energy of $\sqrt s=13$ 
TeV. Proposals are afoot to build accelerators that could extend the range to an ambitious 
goal of 100 TeV.  It has been pointed out recently that there are other avenues to access high
scales and one of these is via sensitive measurement of the EDM of elementary particles, i.e., 
of leptons and of quarks. In this work we focus on the EDM of the electron as it is the most
stringently constrained of the EDMs.  
In this analysis we have used the current experimental limits on the EDM of the electron to 
explore in a quantitative fashion the scale of the slepton masses that the electron 
EDM can explore within MSSM.
It is found that the current constraints allow one to explore a wide scale
of slepton masses from  few TeV to a PeV and beyond. 
Further, we have extended the analysis to include a vector like lepton generation
and allowing for small mixings between the three ordinary generations and the vector like generation.
Here in addition to the supersymmetric contribution involving the exchange of the 
charginos and the neutralinos, one has in addition a  contribution arising from the exchange of the $W$ and of the $Z$ bosons.
Unlike the chargino and the neutralino contribution which is sensitive to the slepton masses,
the $W$ and $Z$ contribution is independent of them. Thus the interference between the 
supersymmetric and the non-supersymmetric contribution produces a remarkable phenomenon
where the EDM first falls and then turns around and rises again as the common scalar mass $m_0$ increases.
This is easily understood by noting that the destructive interference between the supersymmetric 
and the non-supersymmetric contribution leads first to a cancellation between the two but as 
the supersymmetric contribution dies out with increasing $m_0$ the non-supersymmetric 
contribution becomes dominant and controls the EDM. Thus in this case EDM could be substantial
even when $m_0$ lies in the several PeV region. 
 In the future, the EDM of the electron will be constrained even more stringently 
by a factor of ten or more. Such a more stringent constraint will allow one to explore
even a larger range in the slepton masses.
Finally we note that a large
SUSY sfermion scale in the PeV region would automatically relieve the tension on flavor changing
neutral current problem and on too rapid a proton decay in supersymmetric 
grand unified theories ~\cite{Ibrahim:2000tx}.\\

\noindent
{\em Acknowledgments}:
This research was supported in part by the NSF Grant PHY-1314774,
XSEDE- TG-PHY110015, and NERSC-DE-AC02-05CH1123.\\

\noindent
{{\bf Appendix A: The MSSM Extension with a vector leptonic multiplet\label{sec6}}}

{ In \cref{sec3} we extended MSSM to include a vector like  generation. Here we provide 
further details of this extension. 
A vectorlike multiplet consists of an  ordinary fourth generation of
  leptons, quarks and their mirrors. A vector like generation is anomaly free and thus inclusion of it respects the
  good properties of a gauge theory. 
} Vector like multiplets arise in a variety of unified models~\cite{guts} some of which could be low lying.
They have been used recently in a variety of analyses
\cite{Babu:2008ge,Liu:2009cc,Martin:2009bg,Aboubrahim:2013yfa,Aboubrahim:2013gfa,Ibrahim:2008gg,Ibrahim:2010va,Ibrahim:2011im,Ibrahim:2010hv,Ibrahim:2009uv}.
In the analysis below we will assume an extended MSSM with just one  vector mulitplet.
 Before proceeding further we
define the notation and give a very brief description of the extended model and  a more detailed
description can be found in the previous works mentioned above. Thus the extended MSSM 
contains a vectorlike multiplet. To fix notation the three generations of leptons are denoted by

{
\begin{align}
\psi_{iL}\equiv   
 \left(\begin{matrix} \nu_{i L}\cr
 ~{l}_{iL}  \end{matrix} \right)  \sim (1,2,- \frac{1}{2}) \ ;  ~~ ~l^c_{iL}\sim (1,1,1)\ ;
 ~~~ \nu^c_{i L}\sim (1,1,0)\ ;
  ~~~i=1,2,3
\label{2}
\end{align}
}
where the properties  under $SU(3)_C\times SU(2)_L\times U(1)_Y$ are also exhibited.
The last entry in the braces such as $(1,2, -1/2)$ is  
  the value of the hypercharge
 $Y$ defined so that $Q=T_3+ Y$.  These leptons have $V-A$ interactions.
We can now add a vectorlike multiplet where we have a fourth family of leptons with $V-A$ interactions
whose transformations can be gotten from Eq.(\ref{2}) by letting {$i$ run from 1 to 4.}
A vectorlike lepton multiplet also has  mirrors and so we consider these mirror
leptons which have $V+A$ interactions.  {The quantum numbers of the mirrors} are given by

{
\begin{align}
\chi^c\equiv
 \left(\begin{matrix} E_{ L}^c \cr
 N_L^c\end{matrix}\right)  \sim (1,2,\frac{1}{2})\ ; 
~~  E_L \sim  (1,1,-1)\ ;  ~~   N_L \sim (1,1,0).
\label{3}
\end{align}
}

Interesting new physics arises when we allow mixings of the vectorlike generation with
the three ordinary generations.  Here we focus on the mixing of the mirrors in the vectorlike 
generation with the three generations.
Thus the  superpotential of the model allowing for the mixings
among the three ordinary generations and the vectorlike generation is given by

\begin{align}
W&= -\mu \epsilon_{ij} \hat H_1^i \hat H_2^j+\epsilon_{ij}  [f_{1}  \hat H_1^{i} \hat \psi_L ^{j}\hat \tau^c_L
 +f_{1}'  \hat H_2^{j} \hat \psi_L ^{i} \hat \nu^c_{\tau L}
+f_{2}  \hat H_1^{i} \hat \chi^c{^{j}}\hat N_{L}
 +f_{2}'  H_2^{j} \hat \chi^c{^{i}} \hat E_{ L} \nonumber \\
&+ h_{1}  H_1^{i} \hat\psi_{\mu L} ^{j}\hat\mu^c_L
 +h_{1}'  H_2^{j} \hat\psi_{\mu L} ^{i} \hat\nu^c_{\mu L}
+ h_{2}  H_1^{i} \hat\psi_{e L} ^{j}\hat e^c_L
 +h_{2}'  H_2^{j} \hat\psi_{e L} ^{i} \hat\nu^c_{e L}] \nonumber \\
&+ f_{3} \epsilon_{ij}  \hat\chi^c{^{i}}\hat\psi_L^{j}
 + f_{3}' \epsilon_{ij}  \hat\chi^c{^{i}}\hat\psi_{\mu L}^{j}
 + f_{4} \hat\tau^c_L \hat E_{ L}  +  f_{5} \hat\nu^c_{\tau L} \hat N_{L}
 + f_{4}' \hat\mu^c_L \hat E_{ L}  +  f_{5}' \hat\nu^c_{\mu L} \hat N_{L} \nonumber \\
&+ f_{3}'' \epsilon_{ij}  \hat\chi^c{^{i}}\hat\psi_{e L}^{j}
 + f_{4}'' \hat e^c_L \hat E_{ L}  +  f_{5}'' \hat\nu^c_{e L} \hat N_{L}\ ,
 \label{5}
\end{align}
where  $\hat ~$ implies superfields,  $\hat\psi_L$ stands for $\hat\psi_{3L}$, $\hat\psi_{\mu L}$ stands for $\hat\psi_{2L}$
and  $\hat\psi_{e L}$ stands for $\hat\psi_{1L}$.
The mass terms for the neutrinos, mirror neutrinos,  leptons and  mirror leptons arise from the term
\beq
{\cal{L}}=-\frac{1}{2}\frac{\partial ^2 W}{\partial{A_i}\partial{A_j}}\psi_ i \psi_ j+\text{H.c.}
\label{6}
\eeq
where $\psi$ and $A$ stand for generic two-component fermion and scalar fields.
After spontaneous breaking of the electroweak symmetry, ($\langle H_1^1 \rangle=v_1/\sqrt{2} $ and $\langle H_2^2\rangle=v_2/\sqrt{2}$),
we have the following set of mass terms written in the 4-component spinor notation
so that
\beq
-{\cal L}_m= \bar\xi_R^T (M_f) \xi_L +\bar\eta_R^T(M_{\ell}) \eta_L +\text{H.c.},
\eeq
where the basis vectors in which the mass matrix is written is given by
\begin{gather}
\bar\xi_R^T= \left(\begin{matrix}\bar \nu_{\tau R} & \bar N_R & \bar \nu_{\mu R}
&\bar \nu_{e R} \end{matrix}\right),\nonumber\\
\xi_L^T= \left(\begin{matrix} \nu_{\tau L} &  N_L &  \nu_{\mu L}
& \nu_{e L} \end{matrix}\right) \ ,\nonumber\\
\bar\eta_R^T= \left(\begin{matrix}\bar{\tau_ R} & \bar E_R & \bar{\mu_ R}
&\bar{e_ R} \end{matrix}\right),\nonumber\\
\eta_L^T= \left(\begin{matrix} {\tau_ L} &  E_L &  {\mu_ L}
& {e_ L} \end{matrix}\right) \ ,
\end{gather}
and the mass matrix $M_f$ is given by

\beqn
M_f=
 \left(\begin{matrix} f'_1 v_2/\sqrt{2} & f_5 & 0 & 0 \cr
 -f_3 & f_2 v_1/\sqrt{2} & -f_3' & -f_3'' \cr
0&f_5'&h_1' v_2/\sqrt{2} & 0 \cr
0 & f_5'' & 0 & h_2' v_2/\sqrt{2}\end{matrix} \right)\ .
\label{7aa}
\eeqn
We define the matrix element $(22)$ of the mass matrix as $m_N$ so that 
\beqn
m_N= f_2 v_1/\sqrt 2.
\eeqn 
The mass matrix is not hermitian and thus one needs bi-unitary transformations to diagonalize it.
We define the bi-unitary transformation so that

\beq
D^{\nu \dagger}_R (M_f) D^\nu_L=\text{diag}(m_{\psi_1},m_{\psi_2},m_{\psi_3}, m_{\psi_4} ).
\label{7a}
\eeq
Under the bi-unitary transformations the basis vectors transform so that
\beqn
 \left(\begin{matrix} \nu_{\tau_R}\cr
 N_{ R} \cr
\nu_{\mu_R} \cr
\nu_{e_R} \end{matrix}\right)=D^{\nu}_R \left(\begin{matrix} \psi_{1_R}\cr
 \psi_{2_R}  \cr
\psi_{3_R} \cr
\psi_{4_R}\end{matrix}\right), \  \
\left(\begin{matrix} \nu_{\tau_L}\cr
 N_{ L} \cr
\nu_{\mu_L} \cr
\nu_{e_L}\end{matrix} \right)=D^{\nu}_L \left(\begin{matrix} \psi_{1_L}\cr
 \psi_{2_L} \cr
\psi_{3_L} \cr
\psi_{4_L}\end{matrix}\right) \ .
\label{8}
\eeqn
In \cref{7a}
$\psi_1, \psi_2, \psi_3, \psi_4$ are the mass eigenstates for the neutrinos,
where in the limit of no mixing
we identify $\psi_1$ as the light tau neutrino, $\psi_2$ as the
heavier mass eigen state,  $\psi_3$ as the muon neutrino and $\psi_4$ as the electron neutrino.
A similar analysis goes to the lepton mass matrix $M_\ell$ where
\beqn
M_\ell=
 \left(\begin{matrix} f_1 v_1/\sqrt{2} & f_4 & 0 & 0 \cr
 f_3 & f'_2 v_2/\sqrt{2} & f_3' & f_3'' \cr
0&f_4'&h_1 v_1/\sqrt{2} & 0 \cr
0 & f_4'' & 0 & h_2 v_1/\sqrt{2}\end{matrix} \right)\ .
\label{7bb}
\eeqn
In general $f_3, f_4, f_5, f_3', f_4',f_5',  f_3'', f_4'',f_5''$ can be complex and we define their phases
so that

\beqn
f_k= |f_k| e^{i\chi_k}, ~~f_k'= |f_k'| e^{i\chi_k'}, ~~~f_k''= |f_k''| e^{i\chi_k''}\ ;  k=3,4,5\ . 
\eeqn

We introduce now the mass parameter $m_E$ defined by the (22) element of the mass matrix above so that
\beqn
m_E=  f_2' v_2/\sqrt 2.
\eeqn
  Next we  consider  the mixing of the charged sleptons and the charged mirror sleptons.
The mass squared  matrix of the slepton - mirror slepton comes from three sources:  the F term, the
D term of the potential and the soft {SUSY} breaking terms.
Using the  superpotential of \cref{5} the mass terms arising from it
after the breaking of  the electroweak symmetry are given by
the Lagrangian
\beq
{\cal L}= {\cal L}_F +{\cal L}_D + {\cal L}_{\rm soft}\ ,
\eeq
where   $ {\cal L}_F$ is deduced from \cref{5} and is given in \cite{Ibrahim:2012ds}, while the ${\cal L}_D$ is given by
\begin{align}
-{\cal L}_D&=\frac{1}{2} m^2_Z \cos^2\theta_W \cos 2\beta \{\tilde \nu_{\tau L} \tilde \nu^*_{\tau L} -\tilde \tau_L \tilde \tau^*_L
+\tilde \nu_{\mu L} \tilde \nu^*_{\mu L} -\tilde \mu_L \tilde \mu^*_L
+\tilde \nu_{e L} \tilde \nu^*_{e L} -\tilde e_L \tilde e^*_L \nonumber \\
&+\tilde E_R \tilde E^*_R -\tilde N_R \tilde N^*_R\}
+\frac{1}{2} m^2_Z \sin^2\theta_W \cos 2\beta \{\tilde \nu_{\tau L} \tilde \nu^*_{\tau L}
 +\tilde \tau_L \tilde \tau^*_L
+\tilde \nu_{\mu L} \tilde \nu^*_{\mu L} +\tilde \mu_L \tilde \mu^*_L \nonumber \\
&+\tilde \nu_{e L} \tilde \nu^*_{e L} +\tilde e_L \tilde e^*_L
-\tilde E_R \tilde E^*_R -\tilde N_R \tilde N^*_R +2 \tilde E_L \tilde E^*_L -2 \tilde \tau_R \tilde \tau^*_R
-2 \tilde \mu_R \tilde \mu^*_R -2 \tilde e_R \tilde e^*_R
\}.
\label{12}
\end{align}
For ${\cal L}_{\rm soft}$ we assume the following form
\begin{align}
-{\cal L}_{\text{soft}}&= M^2_{\tilde \tau L} \tilde \psi^{i*}_{\tau L} \tilde \psi^i_{\tau L}
+ M^2_{\tilde \chi} \tilde \chi^{ci*} \tilde \chi^{ci}
+ M^2_{\tilde \mu L} \tilde \psi^{i*}_{\mu L} \tilde \psi^i_{\mu L}
+M^2_{\tilde e L} \tilde \psi^{i*}_{e L} \tilde \psi^i_{e L}
+ M^2_{\tilde \nu_\tau} \tilde \nu^{c*}_{\tau L} \tilde \nu^c_{\tau L}
 + M^2_{\tilde \nu_\mu} \tilde \nu^{c*}_{\mu L} \tilde \nu^c_{\mu L} \nonumber \\
&+ M^2_{\tilde \nu_e} \tilde \nu^{c*}_{e L} \tilde \nu^c_{e L}
+ M^2_{\tilde \tau} \tilde \tau^{c*}_L \tilde \tau^c_L
+ M^2_{\tilde \mu} \tilde \mu^{c*}_L \tilde \mu^c_L
+ M^2_{\tilde e} \tilde e^{c*}_L \tilde e^c_L
+ M^2_{\tilde E} \tilde E^*_L \tilde E_L
 +  M^2_{\tilde N} \tilde N^*_L \tilde N_L \nonumber \\
&+\epsilon_{ij} \{f_1 A_{\tau} H^i_1 \tilde \psi^j_{\tau L} \tilde \tau^c_L
-f'_1 A_{\nu_\tau} H^i_2 \tilde \psi ^j_{\tau L} \tilde \nu^c_{\tau L}
+h_1 A_{\mu} H^i_1 \tilde \psi^j_{\mu L} \tilde \mu^c_L
-h'_1 A_{\nu_\mu} H^i_2 \tilde \psi ^j_{\mu L} \tilde \nu^c_{\mu L} \nonumber \\
&+h_2 A_{e} H^i_1 \tilde \psi^j_{e L} \tilde e^c_L
-h'_2 A_{\nu_e} H^i_2 \tilde \psi ^j_{e L} \tilde \nu^c_{e L}
+f_2 A_N H^i_1 \tilde \chi^{cj} \tilde N_L
-f'_2 A_E H^i_2 \tilde \chi^{cj} \tilde E_L +\text{H.c.}\}\ .
\label{13}
\end{align}
Here $M_{\tilde e L}, M_{\tilde \nu_e}$ etc are the soft masses and $A_e, A_{\nu_e}$ etc are the trilinear couplings.
The trilinear couplings are complex  and we define their phases so that 
\begin{gather}
A_e= |A_e| e^{i \alpha_{A_e}} \  ,
 ~~A_{\nu_e}=  |A_{\nu_e}|
 e^{i\alpha_{A_{\nu_e}}} \ , 
  \cdots \ .
\end{gather}
From these terms we construct the scalar mass$^2$ matrices \cite{Ibrahim:2012ds} 
{which are exhibited in Appendix C}.\\

{As discussed in \cref{sec3} and \cref{sec4}  the inclusion of the vector like generation 
  brings in new phenomena such as exchange contributions
  from the $W$ and $Z$ bosons which are otherwise absent.  Their inclusion 
    gives an important contribution to the EDM since the
  $W$ and the $Z$ boson contribution begins to play a role and leads to constructive and
  destructive interference with the chargino and neutralino exchange contribution.
 A more detailed description of this phenomenon is given in \cref{sec4}.}

\noindent
{{\bf Appendix B: Interactions that enter in the EDM analysis in the MSSM Extension with a Vector like Multiplet
\label{sec7}
}}\\
 In this section we discuss the  interactions in the mass diagonal basis involving charged leptons,
 sneutrinos and charginos.  Thus we have
\begin{align}
-{\cal L}_{\tau-\tilde{\nu}-\chi^{-}} &= \sum_{i=1}^{2}\sum_{j=1}^{8}\bar{\tau}_{\alpha}(C_{\alpha ij}^{L}P_{L}+C_{\alpha ij}^{R}P_{R})\tilde{\chi}^{ci}\tilde{\nu}_{j}+\text{H.c.},
\end{align}
such that
\begin{align}
\begin{split}
C_{\alpha ij}^{L}=&g(-\kappa_{\tau}U^{*}_{i2}D^{\tau*}_{R1\alpha} \tilde{D}^{\nu}_{1j} -\kappa_{\mu}U^{*}_{i2}D^{\tau*}_{R3\alpha}\tilde{D}^{\nu}_{5j}-
\kappa_{e}U^{*}_{i2}D^{\tau*}_{R4\alpha}\tilde{D}^{\nu}_{7j}+U^{*}_{i1}D^{\tau*}_{R2\alpha}\tilde{D}^{\nu}_{4j}-
\kappa_{N}U^{*}_{i2}D^{\tau*}_{R2\alpha}\tilde{D}^{\nu}_{2j})
\end{split} \\~\nonumber\\
\begin{split}
C_{\alpha ij}^{R}=&g(-\kappa_{\nu_{\tau}}V_{i2}D^{\tau*}_{L1\alpha}\tilde{D}^{\nu}_{3j}-\kappa_{\nu_{\mu}}V_{i2}D^{\tau*}_{L3\alpha}\tilde{D}^{\nu}_{6j}-
\kappa_{\nu_{e}}V_{i2}D^{\tau*}_{L4\alpha}\tilde{D}^{\nu}_{8j}+V_{i1}D^{\tau*}_{L1\alpha}\tilde{D}^{\nu}_{1j}+V_{i1}D^{\tau*}_{L3\alpha}\tilde{D}^{\nu}_{5j}\\
&+V_{i1}D^{\tau*}_{L4\alpha}\tilde{D}^{\nu}_{7j}-\kappa_{E}V_{i2}D^{\tau*}_{L2\alpha}\tilde{D}^{\nu}_{4j}),
\end{split}
\end{align}
with
\begin{align}
(\kappa_{N},\kappa_{\tau},\kappa_{\mu},\kappa_{e})&=\frac{(m_{N},m_{\tau},m_{\mu},m_{e})}{\sqrt{2}m_{W}\cos\beta} , \\~\nonumber\\
(\kappa_{E},\kappa_{\nu_{\tau}},\kappa_{\nu_{\mu}},\kappa_{\nu_{e}})&=\frac{(m_{E},m_{\nu_{\tau}},m_{\nu_{\mu}},m_{\nu_{e}})}{\sqrt{2}m_{W}\sin\beta} .
\end{align}
   We now  discuss the  interactions in the mass diagonal basis involving charged leptons,
 sleptons and neutralinos.  Thus we have

\begin{align}
-{\cal L}_{\tau-\tilde{\tau}-\chi^{0}} &= \sum_{i=1}^{4}\sum_{j=1}^{8}\bar{\tau}_{\alpha}(C_{\alpha ij}^{'L}P_{L}+C_{\alpha ij}^{'R}P_{R})\tilde{\chi}^{0}_{i}\tilde{\tau}_{j}+\text{H.c.},
\end{align}
such that
\begin{align}
C_{\alpha ij}^{'L}=&\sqrt{2}(\alpha_{\tau i}D^{\tau *}_{R1\alpha}\tilde{D}^{\tau}_{1j}-\delta_{E i}D^{\tau *}_{R2\alpha}\tilde{D}^{\tau}_{2j}-
\gamma_{\tau i}D^{\tau *}_{R1\alpha}\tilde{D}^{\tau}_{3j}+\beta_{E i}D^{\tau *}_{R2\alpha}\tilde{D}^{\tau}_{4j}
+\alpha_{\mu i}D^{\tau *}_{R3\alpha}\tilde{D}^{\tau}_{5j}-\gamma_{\mu i}D^{\tau *}_{R3\alpha}\tilde{D}^{\tau}_{6j} \nonumber\\
&+\alpha_{e i}D^{\tau *}_{R4\alpha}\tilde{D}^{\tau}_{7j}-\gamma_{e i}D^{\tau *}_{R4\alpha}\tilde{D}^{\tau}_{8j})
\end{align}
\begin{align}
C_{\alpha ij}^{'R}=&\sqrt{2}(\beta_{\tau i}D^{\tau *}_{L1\alpha}\tilde{D}^{\tau}_{1j}-\gamma_{E i}D^{\tau *}_{L2\alpha}\tilde{D}^{\tau}_{2j}-
\delta_{\tau i}D^{\tau *}_{L1\alpha}\tilde{D}^{\tau}_{3j}+\alpha_{E i}D^{\tau *}_{L2\alpha}\tilde{D}^{\tau}_{4j}
+\beta_{\mu i}D^{\tau *}_{L3\alpha}\tilde{D}^{\tau}_{5j}-\delta_{\mu i}D^{\tau *}_{L3\alpha}\tilde{D}^{\tau}_{6j}      \nonumber\\
&+\beta_{e i}D^{\tau *}_{L4\alpha}\tilde{D}^{\tau}_{7j}-\delta_{e i}D^{\tau *}_{L4\alpha}\tilde{D}^{\tau}_{8j}),
\end{align}
where

\begin{align}
\alpha_{E i}&=\frac{gm_{E}X^{*}_{4i}}{2m_{W}\sin\beta} \ ;  && \beta_{E i}=eX'_{1i}+\frac{g}{\cos\theta_{W}}X'_{2i}\left(\frac{1}{2}-\sin^{2}\theta_{W}\right) \\
\gamma_{E i}&=eX^{'*}_{1i}-\frac{g\sin^{2}\theta_{W}}{\cos\theta_{W}}X^{'*}_{2i} \  ;  && \delta_{E i}=-\frac{gm_{E}X_{4i}}{2m_{W}\sin\beta}
\end{align}

and
\begin{align}
\alpha_{\tau i}&=\frac{gm_{\tau}X_{3i}}{2m_{W}\cos\beta} \ ;  && \alpha_{\mu i}=\frac{gm_{\mu}X_{3i}}{2m_{W}\cos\beta} \ ; && \alpha_{e i}=\frac{gm_{e}X_{3i}}{2m_{W}\cos\beta}  \\
\delta_{\tau i}&=-\frac{gm_{\tau}X^{*}_{3i}}{2m_{W}\cos\beta} \ ; && \delta_{\mu i}=-\frac{gm_{\mu}X^{*}_{3i}}{2m_{W}\cos\beta} \ ; && \delta_{e i}=-\frac{gm_{e}X^{*}_{3i}}{2m_{W}\cos\beta}
\end{align}
{and where }

\begin{align}
\beta_{\tau i}=\beta_{\mu i}=\beta_{e i}&=-eX^{'*}_{1i}+\frac{g}{\cos\theta_{W}}X^{'*}_{2i}\left(-\frac{1}{2}+\sin^{2}\theta_{W}\right)  \\
\gamma_{\tau i}=\gamma_{\mu i}=\gamma_{e i}&=-eX'_{1i}+\frac{g\sin^{2}\theta_{W}}{\cos\theta_{W}}X'_{2i}
\end{align}
Here $X'$ are defined by

\begin{align}
X'_{1i}&=X_{1i}\cos\theta_{W}+X_{2i}\sin\theta_{W}  \\
X'_{2i}&=-X_ {1i}\sin\theta_{W}+X_{2i}\cos\theta_{W}
\end{align}
where $X$ diagonalizes the neutralino mass matrix and is defined by Eq.(\ref{2.8}).\\

In addition to the computation of the supersymmetric loop diagrams, we compute the contributions
arising from the exchange of the W and $Z$ bosons and the leptons and the mirror leptons in the
loops. The relevant interactions needed are given below. For the $W$ boson exchange the
interactions that enter are given by

\begin{align}
-{\cal L}_{\tau W\psi} &= W^{\dagger}_{\rho}\sum_{i=1}^{4}\sum_{\alpha=1}^{4}\bar{\psi}_{i}\gamma^{\rho}[C_{L_{i\alpha}}^W P_L + C_{R_{i\alpha}}^W P_R]\tau_{\alpha}+\text{H.c.}
\end{align}

where
\beqn
C_{L_{i\alpha}}^W= \frac{g}{\sqrt{2}} [D^{\nu*}_{L1i}D^{\tau}_{L1\alpha}+
D^{\nu*}_{L3i}D^{\tau}_{L3\alpha}+D^{\nu*}_{L4i}D^{\tau}_{L4\alpha}]  \\
C_{R_{i\alpha}}^W= \frac{g}{\sqrt{2}}[D^{\nu*}_{R2i}D^{\tau}_{R2\alpha}]
\eeqn
For the $Z$ boson exchange the interactions that enter are given by

\beqn
-{\cal L}_{\tau\tau Z} &= Z_{\rho}\sum_{\alpha=1}^{4}\sum_{\beta=1}^{4}\bar{\tau}_{\alpha}\gamma^{\rho}[C_{L_{\alpha \beta}}^Z P_L + C_{R_{\alpha \beta}}^Z P_R]\tau_{\beta}
\eeqn
 where
\beqn
C_{L_{\alpha \beta}}^Z=\frac{g}{\cos\theta_{W}} [x(D_{L\alpha 1}^{\tau\dag}D_{L1\beta}^{\tau}+D_{L\alpha 2}^{\tau\dag}D_{L2\beta}^{\tau}+D_{L\alpha 3}^{\tau\dag}D_{L3\beta}^{\tau}+D_{L\alpha 4}^{\tau\dag}D_{L4\beta}^{\tau})\nonumber\\
-\frac{1}{2}(D_{L\alpha 1}^{\tau\dag}D_{L1\beta}^{\tau}+D_{L\alpha 3}^{\tau\dag}D_{L3\beta}^{\tau}+D_{L\alpha 4}^{\tau\dag}D_{L4\beta}^{\tau})]
\eeqn
and
\beqn
C_{R_{\alpha \beta}}^Z=\frac{g}{\cos\theta_{W}} [x(D_{R\alpha 1}^{\tau\dag}D_{R1\beta}^{\tau}+D_{R\alpha 2}^{\tau\dag}D_{R2\beta}^{\tau}+D_{R\alpha 3}^{\tau\dag}D_{R3\beta}^{\tau}+D_{R\alpha 4}^{\tau\dag}D_{R4\beta}^{\tau})\nonumber\\
-\frac{1}{2}(D_{R\alpha 2}^{\tau\dag}
D_{R 2\beta }^{\tau}
 )]
\eeqn
where $x=\sin^{2}\theta_{W}$.\\

\noindent
{{\bf Appendix C : The scalar mass  squared matrices
   \label{sec8}}}\\
   For convenience we collect here all the contributions to the scalar mass$^2$ matrices
   arising from the superpotential.  They are given by
\beq
{\cal L}^{\rm mass}_F= {\cal L}_C^{\rm mass} +{\cal L}_N^{\rm mass}\ ,
\eeq
where  ${\cal L}_C^{\rm mass}$ gives the mass terms for the charged sleptons while
$ {\cal L}_N^{mass}$ gives the mass terms for the  sneutrinos. For ${\cal L}_C^{\rm mass}$ we have
\begin{gather}
-{\cal L}_C^{\rm mass} =\left(\frac{v^2_2 |f'_2|^2}{2} +|f_3|^2+|f_3'|^2+|f_3''|^2\right)\tilde E_R \tilde E^*_R
+\left(\frac{v^2_2 |f'_2|^2}{2} +|f_4|^2+|f_4'|^2+|f_4''|^2\right)\tilde E_L \tilde E^*_L\nonumber\\
+\left(\frac{v^2_1 |f_1|^2}{2} +|f_4|^2\right)\tilde \tau_R \tilde \tau^*_R
+\left(\frac{v^2_1 |f_1|^2}{2} +|f_3|^2\right)\tilde \tau_L \tilde \tau^*_L
+\left(\frac{v^2_1 |h_1|^2}{2} +|f_4'|^2\right)\tilde \mu_R \tilde \mu^*_R\nonumber\\
+\left(\frac{v^2_1 |h_1|^2}{2} +|f_3'|^2\right)\tilde \mu_L \tilde \mu^*_L
+\left(\frac{v^2_1 |h_2|^2}{2} +|f_4''|^2\right)\tilde e_R \tilde e^*_R
+\left(\frac{v^2_1 |h_2|^2}{2} +|f_3''|^2\right)\tilde e_L \tilde e^*_L\nonumber\\
+\Bigg\{-\frac{f_1 \mu^* v_2}{\sqrt{2}} \tilde \tau_L \tilde \tau^*_R
-\frac{h_1 \mu^* v_2}{\sqrt{2}} \tilde \mu_L \tilde \mu^*_R
 -\frac{f'_2 \mu^* v_1}{\sqrt{2}} \tilde E_L \tilde E^*_R
+\left(\frac{f'_2 v_2 f^*_3}{\sqrt{2}}  +\frac{f_4 v_1 f^*_1}{\sqrt{2}}\right) \tilde E_L \tilde \tau^*_L\nonumber\\
+\left(\frac{f_4 v_2 f'^*_2}{\sqrt{2}}  +\frac{f_1 v_1 f^*_3}{\sqrt{2}}\right) \tilde E_R \tilde \tau^*_R
+\left(\frac{f'_3 v_2 f'^*_2}{\sqrt{2}}  +\frac{h_1 v_1 f'^*_4}{\sqrt{2}}\right) \tilde E_L \tilde \mu^*_L
+\left(\frac{f'_2 v_2 f'^*_4}{\sqrt{2}}  +\frac{f'_3 v_1 h^*_1}{\sqrt{2}}\right) \tilde E_R \tilde \mu^*_R\nonumber\\
+\left(\frac{f''^*_3 v_2 f'_2}{\sqrt{2}}  +\frac{f''_4 v_1 h^*_2}{\sqrt{2}}\right) \tilde E_L \tilde e^*_L
+\left(\frac{f''_4 v_2 f'^*_2}{\sqrt{2}}  +\frac{f''^*_3 v_1 h^*_2}{\sqrt{2}}\right) \tilde E_R \tilde e^*_R
+f'_3 f^*_3 \tilde \mu_L \tilde \tau^*_L +f_4 f'^*_4 \tilde \mu_R \tilde \tau^*_R\nonumber\\
+f_4 f''^*_4 \tilde {e}_R \tilde{\tau}^*_R
+f''_3 f^*_3 \tilde {e}_L \tilde{\tau}^*_L
+f''_3 f'^*_3 \tilde {e}_L \tilde{\mu}^*_L
+f'_4 f''^*_4 \tilde {e}_R \tilde{\mu}^*_R
-\frac{h_2 \mu^* v_2}{\sqrt{2}} \tilde{e}_L \tilde{e}^*_R
+H.c. \Bigg\}
\end{gather}
We define the scalar mass squared   matrix $M^2_{\tilde \tau}$  in the basis $(\tilde  \tau_L, \tilde E_L, \tilde \tau_R,
\tilde E_R, \tilde \mu_L, \tilde \mu_R, \tilde e_L, \tilde e_R)$. We  label the matrix  elements of these as $(M^2_{\tilde \tau})_{ij}= M^2_{ij}$ where the elements of the matrix are given by
\begin{align}
M^2_{11}&=\tilde M^2_{\tau L} +\frac{v^2_1|f_1|^2}{2} +|f_3|^2 -m^2_Z \cos 2 \beta \left(\frac{1}{2}-\sin^2\theta_W\right), \nonumber\\
M^2_{22}&=\tilde M^2_E +\frac{v^2_2|f'_2|^2}{2}+|f_4|^2 +|f'_4|^2+|f''_4|^2 +m^2_Z \cos 2 \beta \sin^2\theta_W, \nonumber\\
M^2_{33}&=\tilde M^2_{\tau} +\frac{v^2_1|f_1|^2}{2} +|f_4|^2 -m^2_Z \cos 2 \beta \sin^2\theta_W, \nonumber\\
M^2_{44}&=\tilde M^2_{\chi} +\frac{v^2_2|f'_2|^2}{2} +|f_3|^2 +|f'_3|^2+|f''_3|^2 +m^2_Z \cos 2 \beta \left(\frac{1}{2}-\sin^2\theta_W\right), \nonumber
\end{align}
\begin{align}
M^2_{55}&=\tilde M^2_{\mu L} +\frac{v^2_1|h_1|^2}{2} +|f'_3|^2 -m^2_Z \cos 2 \beta \left(\frac{1}{2}-\sin^2\theta_W\right), \nonumber\\
M^2_{66}&=\tilde M^2_{\mu} +\frac{v^2_1|h_1|^2}{2}+|f'_4|^2 -m^2_Z \cos 2 \beta \sin^2\theta_W, \nonumber\\
M^2_{77}&=\tilde M^2_{e L} +\frac{v^2_1|h_2|^2}{2}+|f''_3|^2 -m^2_Z \cos 2 \beta \left(\frac{1}{2}-\sin^2\theta_W\right), \nonumber\\
M^2_{88}&=\tilde M^2_{e} +\frac{v^2_1|h_2|^2}{2}+|f''_4|^2 -m^2_Z \cos 2 \beta \sin^2\theta_W\ . \nonumber
\end{align}

\begin{align}
M^2_{12}&=M^{2*}_{21}=\frac{ v_2 f'_2f^*_3}{\sqrt{2}} +\frac{ v_1 f_4 f^*_1}{\sqrt{2}} ,
M^2_{13}=M^{2*}_{31}=\frac{f^*_1}{\sqrt{2}}(v_1 A^*_{\tau} -\mu v_2),
M^2_{14}=M^{2*}_{41}=0,\nonumber\\
 M^2_{15} &=M^{2*}_{51}=f'_3 f^*_3,
 M^{2*}_{16}= M^{2*}_{61}=0,  M^{2*}_{17}= M^{2*}_{71}=f''_3 f^*_3,  M^{2*}_{18}= M^{2*}_{81}=0,\nonumber\\
M^2_{23}&=M^{2*}_{32}=0,
M^2_{24}=M^{2*}_{42}=\frac{f'^*_2}{\sqrt{2}}(v_2 A^*_{E} -\mu v_1),  M^2_{25} = M^{2*}_{52}= \frac{ v_2 f'_3f'^*_2}{\sqrt{2}} +\frac{ v_1 h_1 f^*_4}{\sqrt{2}} ,\nonumber\\
 M^2_{26} &=M^{2*}_{62}=0,  M^2_{27} =M^{2*}_{72}=  \frac{ v_2 f''_3f'^*_2}{\sqrt{2}} +\frac{ v_1 h_1 f'^*_4}{\sqrt{2}},  M^2_{28} =M^{2*}_{82}=0, \nonumber\\
M^2_{34}&=M^{2*}_{43}= \frac{ v_2 f_4 f'^*_2}{\sqrt{2}} +\frac{ v_1 f_1 f^*_3}{\sqrt{2}}, M^2_{35} =M^{2*}_{53} =0, M^2_{36} =M^{2*}_{63}=f_4 f'^*_4,\nonumber\\
 M^2_{37} &=M^{2*}_{73} =0,  M^2_{38} =M^{2*}_{83} =f_4 f''^*_4,\nonumber\\
M^2_{45}&=M^{2*}_{54}=0, M^2_{46}=M^{2*}_{64}=\frac{ v_2 f'_2 f'^*_4}{\sqrt{2}} +\frac{ v_1 f'_3 h^*_1}{\sqrt{2}}, \nonumber\\
 M^2_{47} &=M^{2*}_{74}=0,  M^2_{48} =M^{2*}_{84}=  \frac{ v_2 f'_2f''^*_4}{\sqrt{2}} +\frac{ v_1 f''_3 h^*_2}{\sqrt{2}},\nonumber\\
M^2_{56}&=M^{2*}_{65}=\frac{h^*_1}{\sqrt{2}}(v_1 A^*_{\mu} -\mu v_2),
 M^2_{57} =M^{2*}_{75}=f''_3 f'^*_3,  \nonumber\\
 M^2_{58} &=M^{2*}_{85}=0,  M^2_{67} =M^{2*}_{76}=0,\nonumber\\
 M^2_{68} &=M^{2*}_{86}=f'_4 f''^*_4,  M^2_{78}=M^{2*}_{87}=\frac{h^*_2}{\sqrt{2}}(v_1 A^*_{e} -\mu v_2)\ . \nonumber
\label{14}
\end{align}

We can diagonalize this hermitian mass squared  matrix  by the
 unitary transformation
\begin{gather}
 \tilde D^{\tau \dagger} M^2_{\tilde \tau} \tilde D^{\tau} = diag (M^2_{\tilde \tau_1},
M^2_{\tilde \tau_2}, M^2_{\tilde \tau_3},  M^2_{\tilde \tau_4},  M^2_{\tilde \tau_5},  M^2_{\tilde \tau_6},  M^2_{\tilde \tau_7},  M^2_{\tilde \tau_8} )\ .
\end{gather}


For ${\cal L}_N^{\rm mass}$ we have
\begin{multline}
-{\cal L}_N^{\rm mass}=
\left(\frac{v^2_1 |f_2|^2}{2}
 +|f_3|^2+|f_3'|^2+|f_3''|^2\right)\tilde N_R \tilde N^*_R\\
 +\left(\frac{v^2_1 |f_2|^2}{2}+|f_5|^2+|f_5'|^2+|f_5''|^2\right)\tilde N_L \tilde N^*_L
+\left(\frac{v^2_2 |f'_1|^2}{2}+|f_5|^2\right)\tilde \nu_{\tau R} \tilde \nu^*_{\tau R}\\
+\left(\frac{v^2_2 |f'_1|^2}{2}
+|f_3|^2\right)\tilde \nu_{\tau L} \tilde \nu^*_{\tau L}
+\left(\frac{v^2_2 |h'_1|^2}{2}
+|f_3'|^2\right)\tilde \nu_{\mu L} \tilde \nu^*_{\mu L}
+\left(\frac{v^2_2 |h'_1|^2}{2}
+|f_5'|^2\right)\tilde \nu_{\mu R} \tilde \nu^*_{\mu R}\nonumber\\
+\left(\frac{v^2_2 |h'_2|^2}{2}
+|f_3''|^2\right)\tilde \nu_{e L} \tilde \nu^*_{e L}
+\left(\frac{v^2_2 |h'_2|^2}{2}
+|f_5''|^2\right)\tilde \nu_{e R} \tilde \nu^*_{e R}\nonumber\\
+\Bigg\{ -\frac{f_2 \mu^* v_2}{\sqrt{2}} \tilde N_L \tilde N^*_R
-\frac{f'_1 \mu^* v_1}{\sqrt{2}} \tilde \nu_{\tau L} \tilde \nu^*_{\tau R}
-\frac{h'_1 \mu^* v_1}{\sqrt{2}} \tilde \nu_{\mu L} \tilde \nu^*_{\mu R}
+\left(\frac{f_5 v_2 f'^*_1}{\sqrt{2}}  -\frac{f_2 v_1 f^*_3}{\sqrt{2}}\right) \tilde N_L \tilde \nu^*_{\tau L}\nonumber\\
+\left(\frac{f_5 v_1 f^*_2}{\sqrt{2}}  -\frac{f'_1 v_2 f^*_3}{\sqrt{2}}\right) \tilde N_R \tilde \nu^*_{\tau R}
+\left(\frac{h'_1 v_2 f'^*_5}{\sqrt{2}}  -\frac{f'_3 v_1 f^*_2}{\sqrt{2}}\right) \tilde N_L \tilde \nu^*_{\mu L}
+\left(\frac{f''_5 v_1 f^*_2}{\sqrt{2}}  -\frac{f''^*_3 v_2 h'_2}{\sqrt{2}}\right) \tilde N_R \tilde \nu^*_{e R}\nonumber\\
+\left(\frac{h'^*_2 v_2 f''_5}{\sqrt{2}}  -\frac{f''^*_3 v_1 f_2}{\sqrt{2}}\right) \tilde N_L \tilde \nu^*_{e L}
+\left(\frac{f'_5 v_1 f^*_2}{\sqrt{2}}  -\frac{h'_1 v_2 f'^*_3}{\sqrt{2}}\right) \tilde N_R \tilde \nu^*_{\mu R}\nonumber\\
+f'_3 f^*_3 \tilde \nu_{\mu L} \tilde \nu_{\tau^*_L} +f_5 f'^*_5 \tilde \nu_{\mu R} \tilde \nu^*_{\tau R}
-\frac{h'_2 \mu^* v_1}{\sqrt{2}} \tilde{\nu}_{e L} \tilde{\nu}^*_{e R}\\
+f''_3 f^*_3   \tilde{\nu}_{e L} \tilde{\nu}^*_{\tau L}
+f_5 f''^*_5   \tilde{\nu}_{e R} \tilde{\nu}^*_{\tau R}
+f''_3 f'^*_3   \tilde{\nu}_{e L} \tilde{\nu}^*_{\mu L}
+f'_5 f''^*_5   \tilde{\nu}_{e R} \tilde{\nu}^*_{\mu R}
+H.c. \Bigg\}.
\label{11b}
\end{multline}

 Next we write the   mass$^2$  matrix in the sneutrino sector the basis $(\tilde  \nu_{\tau L}, \tilde N_L,$
$ \tilde \nu_{\tau R}, \tilde N_R, \tilde  \nu_{\mu L},\tilde \nu_{\mu R}, \tilde \nu_{e L}, \tilde \nu_{e R} )$.
 Thus here we denote the sneutrino mass$^2$ matrix in the form
$(M^2_{\tilde\nu})_{ij}=m^2_{ij}$ where

\begin{align}
m^2_{11}&=\tilde M^2_{\tau L} +m^2_{\nu_\tau} +|f_3|^2 +\frac{1}{2}m^2_Z \cos 2 \beta,  \nonumber\\
m^2_{22}&=\tilde M^2_N +m^2_{N} +|f_5|^2 +|f'_5|^2+|f''_5|^2, \nonumber\\
m^2_{33}&=\tilde M^2_{\nu_\tau} +m^2_{\nu_\tau} +|f_5|^2,  \nonumber\\
m^2_{44}&=\tilde M^2_{\chi} +m^2_{N} +|f_3|^2 +|f'_3|^2+|f''_3|^2 -\frac{1}{2}m^2_Z \cos 2 \beta, \nonumber\\
m^2_{55}&=\tilde M^2_{\mu L} +m^2_{\nu_\mu} +|f'_3|^2 +\frac{1}{2}m^2_Z \cos 2 \beta,  \nonumber\\
m^2_{66}&=\tilde M^2_{\nu_\mu} +m^2_{\nu_\mu} +|f'_5|^2,  \nonumber\\
m^2_{77}&=\tilde M^2_{e L} +m^2_{\nu_e} +|f''_3|^2+\frac{1}{2}m^2_Z \cos 2 \beta,  \nonumber\\
m^2_{88}&=\tilde M^2_{\nu_e} +m^2_{\nu_e} +|f''_5|^2,  \nonumber
\end{align}

\begin{align}
m^2_{12}&=m^{2*}_{21}=\frac{v_2 f_5 f'^*_1}{\sqrt{2}}-\frac{ v_1 f_2 f^*_3}{\sqrt{2}},
~m^2_{13}=m^{2*}_{31}=\frac{f'^*_1}{\sqrt{2}}(v_2 A^*_{\nu_\tau} -\mu v_1)\nonumber\\
m^2_{14}&=m^{2*}_{41}=0,
~m^2_{15}=m^{2*}_{51}= f'_3 f^*_3, m^2_{16}=m^{2*}_{61}=0,\nonumber\\
m^2_{17}&=m^{2*}_{71}= f''_3 f^*_3, m^2_{18}=m^{2*}_{81}=0,\nonumber\\
m^2_{23}&=m^{2*}_{32}=0,
m^2_{24}=m^{2*}_{42}=\frac{f^*_2}{\sqrt{2}}(v_{1}A^*_N-\mu v_2), \nonumber\\
m^2_{25}&=m^{2*}_{52}=-\frac{v_{1}f^*_2 f'_3}{\sqrt{2}}+\frac{h'_1 v_2 f'^*_5}{\sqrt{2}},\nonumber\\
m^2_{26}&=m^{2*}_{62}=0, m^2_{27}=m^{2*}_{72}=-\frac{v_{1}f^*_2 f''_3}{\sqrt{2}}+\frac{h'_2 v_2 f''^*_5}{\sqrt{2}}
\nonumber\\ 
m^2_{28}&=m^{2*}_{82}=0, m^2_{34}=m^{2*}_{43}=\frac{v_1 f^*_2 f_5}{\sqrt{2}}-\frac{v_2 f'_1 f^*_3}{\sqrt{2}},\nonumber\\
m^2_{35}&=m^{2*}_{53}=0, m^2_{36}=m^{2*}_{63}=f_5 f'^*_5, \nonumber\\
m^2_{37}&=m^{2*}_{73}=0, m^2_{38}=m^{2*}_{83}=f_5 f''^*_5, \nonumber\\
m^2_{45}&=m^{2*}_{54}=0, m^2_{46}=m^{2*}_{64}=-\frac{h'^*_1 v_2 f'_3}{\sqrt{2}}+\frac{v_1 f_2 f'^*_5}{\sqrt{2}}, 
\nonumber\\
m^2_{47}&=m^{2*}_{74}=0, 
m^2_{48}=m^{2*}_{84}=\frac{v_1 f_2 f''^*_5}{\sqrt{2}}-\frac{v_2 h'^*_2 f''_3}{\sqrt{2}},\nonumber\\
 m^2_{56}&=m^{2*}_{65}=\frac{h'^*_1}{\sqrt{2}}(v_2 A^*_{\nu_\mu}-\mu v_1), \nonumber\\
m^2_{57}&=m^{2*}_{75}= f''_3 f'^*_3, m^2_{58}=m^{2*}_{85}=0, \nonumber\\
m^2_{67}&=m^{2*}_{76}=0, m^2_{68}=m^{2*}_{86}= f'_5 f''^*_5, \nonumber\\
m^2_{78}&=m^{2*}_{87}=\frac{h'^*_2}{\sqrt{2}}(v_2 A^*_{\nu_e}-\mu v_1).  
\end{align}

We can diagonalize the sneutrino mass square matrix  by the  unitary transformation 
\begin{equation}
 \tilde D^{\nu\dagger} M^2_{\tilde \nu} \tilde D^{\nu} = \text{diag} (M^2_{\tilde \nu_1}, M^2_{\tilde \nu_2}, M^2_{\tilde \nu_3},  M^2_{\tilde \nu_4},M^2_{\tilde \nu_5},  M^2_{\tilde \nu_6}, M^2_{\tilde \nu_7}, M^2_{\tilde \nu_8})\ .
\end{equation}

\newpage

\end{document}